\documentclass[twocolumn]{aastex63}

\usepackage{rotating}
\usepackage{acronym}
\usepackage{xspace}
\usepackage{multirow}
\usepackage{mathtools}
\usepackage{amsmath}
\usepackage{hyperref}

\usepackage{xcolor}

\graphicspath{{./}{figures/}}

\definecolor{chmagenta}{rgb}{0.54, 0.17, 0.88}

\def\Mc{\ensuremath{\mathcal{M}_\mathrm{c}}\xspace}
\def\chieff{\ensuremath{\chi_\mathrm{eff}}\xspace}
\def\q{\ensuremath{q}\xspace}
\def\z{\ensuremath{z}\xspace}
\def\chib{\ensuremath{\chi_\mathrm{b}}\xspace}
\def\alphaCE{\ensuremath{\alpha_\mathrm{CE}}\xspace}
\def\BF{\ensuremath{\mathcal{B}}\xspace}

\def\Msun{\ensuremath{\mathit{M_\odot}}\xspace}
\def\Zsun{\ensuremath{\mathit{Z_\odot}}\xspace}

\def\BFchifiveCEGC{\ensuremath{40}\xspace}
\def\BFchitwoCEGC{\ensuremath{7}\xspace}
\def\BFchioneCEGC{\ensuremath{4}\xspace}

\def\BetaCETwochannel{\ensuremath{0.89_{-0.22}^{+0.07}}\xspace}
\def\BetaGCTwochannel{\ensuremath{0.11_{-0.07}^{+0.22}}\xspace}
\def\BetaCETwochannelChiFive{\ensuremath{0.26_{-0.23}^{+0.38}}\xspace}
\def\BetaGCTwochannelChiFive{\ensuremath{0.74_{-0.38}^{+0.23}}\xspace}

\def\MaxBetaNinetyCEGC{\ensuremath{0.95}\xspace}
\def\MaxBetaNinetyNineCEGC{\ensuremath{0.97}\xspace}

\def\BetaCE{\ensuremath{0.708_{-0.604}^{+0.193}}\xspace}
\def\BetaCHE{\ensuremath{0.023_{-0.018}^{+0.058}}\xspace}
\def\BetaGC{\ensuremath{0.114_{-0.091}^{+0.298}}\xspace}
\def\BetaNSC{\ensuremath{0.024_{-0.021}^{+0.105}}\xspace}
\def\BetaSMT{\ensuremath{0.100_{-0.090}^{+0.365}}\xspace}

\def\PrecentLevelConstraintsLow{\ensuremath{80\%}\xspace}
\def\PrecentLevelConstraintsHigh{\ensuremath{430\%}\xspace}

\def\BetaCEOnePercent{\ensuremath{2.4\%}\xspace}

\def\PrecentDifferenceCE{\ensuremath{21\%}\xspace}
\def\PrecentDifferenceGC{\ensuremath{5\%}\xspace}
\def\ChiOneBFCompare{\ensuremath{2.2}\xspace}

\def\BetaIncreaseGC{\ensuremath{0.51}\xspace}
\def\BetaIncreaseCE{\ensuremath{0.63}\xspace}

\def\BetaDetCE{\ensuremath{0.08_{-0.07}^{+0.18}}\xspace}
\def\BetaDetCHE{\ensuremath{0.11_{-0.08}^{+0.12}}\xspace}
\def\BetaDetGC{\ensuremath{0.30_{-0.22}^{+0.26}}\xspace}
\def\BetaDetNSC{\ensuremath{0.19_{-0.16}^{+0.26}}\xspace}
\def\BetaDetSMT{\ensuremath{0.26_{-0.24}^{+0.27}}\xspace}

\def\BetaDetectableGCNinetyNine{\ensuremath{2.6\%}\xspace}

\def\BetaDetectableSMTIncrease{\ensuremath{24\%}\xspace}

\def\PosteriorNinetyBmax{\ensuremath{0.88}\xspace}
\def\PosteriorNinetyNineBmax{\ensuremath{0.93}\xspace}
\def\PriorNinetyBmax{\ensuremath{0.62}\xspace}
\def\PriorNinetyNineBmax{\ensuremath{0.79}\xspace}

\def\PosteriorNinetyBmaxDetectable{\ensuremath{0.56}\xspace}
\def\PosteriorNinetyNineBmaxDetectable{\ensuremath{0.69}\xspace}
\def\PriorNinetyBmaxDetectable{\ensuremath{0.75}\xspace}
\def\PriorNinetyNineBmaxDetectable{\ensuremath{0.89}\xspace}
\def\PosteriorNinetyNineBmaxDetectablePercentApprox{\ensuremath{70\%}\xspace}

\def\NOneBetaThreshold{\ensuremath{26\%}\xspace}
\def\NLeqThreeBetaThreshold{\ensuremath{95\%}\xspace}

\def\NGtrOneThresholdDetectable{\ensuremath{99.8\%}\xspace}
\def\NThreeFourThresholdDetectable{\ensuremath{79\%}\xspace}

\def\BFchilowchihigh{\ensuremath{12.9}\xspace}
\def\BFchizerochione{\ensuremath{1.9}\xspace}

\def\BFalphafive{\ensuremath{5}\xspace}
\def\BFalphalowest{\ensuremath{10}\xspace}

\def\BFchizerochifive{\ensuremath{27}\xspace}

\def\BetaField{\ensuremath{0.86_{-0.36}^{+0.11}}\xspace}
\def\BetaDynamical{\ensuremath{0.14_{-0.11}^{+0.36}}\xspace}
\def\BetaFieldChiFive{\ensuremath{0.18_{-0.12}^{+0.27}}\xspace}
\def\BetaDynamicalChiFive{\ensuremath{0.82_{-0.27}^{+0.12}}\xspace}

\def\BetaFieldDet{\ensuremath{0.50_{-0.30}^{+0.25}}\xspace}
\def\BetaDynamicalDet{\ensuremath{0.50_{-0.25}^{+0.30}}\xspace}

\acrodef{GW}{gravitational-wave}
\acrodef{LVC}{LIGO Scientific and Virgo Collaboration}
\acrodef{BBH}{binary black hole}
\acrodef{BH}{black hole}
\acrodef{SNR}{signal-to-noise ratio}
\acrodef{SFR}{star formation rate}
\acrodef{KDE}{kernel density estimate}
\acrodef{BSE}{Binary Stellar Evolution}

\acrodef{CE}{common envelope}
\acrodef{GC}{globular cluster}
\acrodef{NSC}{nuclear star cluster}
\acrodef{AGN}{active galactic nucleus}
\acrodefplural{AGN}[AGNs]{active galactic nuclei}

\acrodef{LIGO}{Laser Interferometer Gravitational-wave Observatory}
\acrodef{NS}{black hole}
\acrodef{BNS}{binary neutron star}
\acrodef{O1}{first observing run}
\acrodef{O2}{second observing run}
\acrodef{O3}{third observing run}
\acrodef{O3a}{first half of the third observing run}
\acrodef{GR}{general relativity}
\acrodef{PSD}{power spectral density}
\acrodef{NSF}{National Science Foundation}

\newcommand{\Northwestern}{\affiliation{Department of Physics and Astronomy, Northwestern University, 2145 Sheridan Road, Evanston, IL 60208, USA}}
\newcommand{\CIERA}{\affiliation{Center for Interdisciplinary Exploration and Research in Astrophysics (CIERA), 1800 Sherman Avenue, Evanston, IL 60201, USA}}
\newcommand{\UChicago}{\affiliation{Enrico Fermi Institute and Kavli Institute for Cosmological Physics, The University of Chicago, 5640 South Ellis Avenue, Chicago, Illinois 60637, USA}}
\newcommand{\Geneva}{\affiliation{Geneva Observatory, University of Geneva, Chemin Pegasi 51, 1290 Versoix, Switzerland}}
\newcommand{\SUPA}{\affiliation{SUPA, School of Physics and Astronomy, University of Glasgow, Glasgow G12 8QQ, UK}}
\newcommand{\Leuven}{\affiliation{Institute of Astrophysics, KU Leuven, Celestijnenlaan 200D, B-3001, Leuven, Belgium}}
\newcommand{\CarnegieMellon}{\affiliation{Department of Physics, Carnegie Mellon University, 5000 Forbes Avenue, Pittsburgh, PA 15213, USA}}
\newcommand{\Cardiff}{\affiliation{Gravity Exploration Institute, School of Physics and Astronomy, Cardiff University, Cardiff, CF24 3AA, United Kingdom}}
\newcommand{\UChicagoTwo}{\affiliation{Department of Physics, Department of Astronomy \& Astrophysics, The University of Chicago, 5640 South Ellis Avenue, Chicago, Illinois 60637, USA}}

\shorttitle{Constraining Origins of BBHs with GWTC-2}
\shortauthors{Zevin et al.}

\begin{document}

\title{One Channel to Rule Them All? \\
Constraining the Origins of Binary Black Holes Using Multiple Formation Pathways}

\author[0000-0002-0147-0835]{Michael\,Zevin}\thanks{NASA Hubble Fellow}\thanks{michaelzevin@uchicago.edu}
\Northwestern
\CIERA
\UChicago

\author[0000-0002-3439-0321]{Simone\,S.\,Bavera}
\Geneva
\CIERA

\author[0000-0003-3870-7215]{Christopher\,P.\,L.\,Berry}
\CIERA
\SUPA

\author[0000-0001-9236-5469]{Vicky\,Kalogera}
\Northwestern
\CIERA

\author[0000-0003-1474-1523]{Tassos\,Fragos}
\Geneva

\author[0000-0002-0338-8181]{Pablo\,Marchant}
\Leuven

\author[0000-0003-4175-8881]{Carl\,L.\,Rodriguez}
\CarnegieMellon

\author[0000-0003-3138-6199]{Fabio\,Antonini}
\Cardiff

\author[0000-0002-0175-5064]{Daniel\,E.\,Holz}
\UChicago
\UChicagoTwo

\author[0000-0002-1128-3662]{Chris\,Pankow}
\CIERA

\begin{abstract}
The second LIGO--Virgo catalog of gravitational-wave (GW) transients has more than quadrupled the observational sample of binary black holes. 
We analyze this catalog using a suite of five state-of-the-art binary black hole population models covering a range of isolated and dynamical formation channels and infer branching fractions between channels as well as constraints on uncertain physical processes that impact the observational properties of mergers. 
Given our set of formation models, we find significant differences between the branching fractions of the underlying and detectable populations, and that the diversity of detections suggests that multiple formation channels are at play. 
A mixture of channels is strongly preferred over any single channel dominating the detected population: an individual channel does not contribute to more than $\simeq \PosteriorNinetyNineBmaxDetectablePercentApprox$ of the observational sample of binary black holes. 
We calculate the preference between the natal spin assumptions and common-envelope efficiencies in our models, favoring natal spins of isolated black holes of $\lesssim 0.1$, and marginally preferring common-envelope efficiencies of $\gtrsim 2.0$ while strongly disfavoring highly inefficient common envelopes. 
We show that it is essential to consider multiple channels when interpreting GW catalogs, as inference on branching fractions and physical prescriptions becomes biased when contributing formation scenarios are not considered or incorrect physical prescriptions are assumed. 
Although our quantitative results can be affected by uncertain assumptions in model predictions, our methodology is capable of including models with updated theoretical considerations and additional formation channels. 
\end{abstract}


\section{Introduction}\label{sec:intro}

In less than five years the field of \ac{GW} astrophysics has evolved from speculating about the properties of compact binary coalescence events to having a substantial population primed for astrophysical inference. 
The recently released catalog of compact binary coalescences (GWTC-2), accumulated by the LIGO and Virgo \ac{GW} detector network~\citep{aLIGO,aVirgo}, has increased the number of confident detections reported by the \ac{LVC} to $50$~\citep{GWTC2}. 
As the endpoint of massive-star evolution, merging double compact objects can encode unique information about their progenitor systems, such as the types of galactic environments they were born in and their formation processes, the complex stellar evolution that persisted throughout their lives, and the physics of the supernovae that marked their deaths~\citep{GW150914_astro,Mandel2018,Vitale2020}. 
From this catalog, the rates of compact binary mergers in the local universe have been significantly constrained, features have been resolved in the \ac{BBH} mass spectrum, and a non-negligible fraction of systems have been found to have spins misaligned relative to the pre-merger orbital angular momentum by more than $90^\circ$~\citep{GWTC2_pops}.

Of the GWTC-2 observations, the vast majority ($46$) are confidently identified as \ac{BBH} mergers~\citep{GWTC2}. 
\acp{BBH} have a disparate array of proposed formation channels. 
The simple picture of two canonical \ac{BBH} formation channels, the isolated evolution of a massive-star binary and dynamical assembly in a dense stellar environment, is now inadequate to capture the breadth of theoretical models proposed for \ac{BBH} mergers. 
Both the isolated evolution and dynamical assembly paradigms have multiple subchannels with differing predictions for mass distributions, spin distributions, and the redshift evolution of \ac{BBH} mergers, each with predicted local merger rates consistent with the empirical rate measured by the \ac{LVC} of $15$--$40~\mathrm{Gpc^{-3}\, yr^{-1}}$~\citep[90\% credible interval;][]{GWTC2_pops}. 

On the isolated evolution side, the standard channel involves a phase of unstable mass transfer following the formation of the first \ac{BH}, initiating a \ac{CE} phase that hardens the binary via drag forces~\citep{Paczynski1976,vandenHeuvel1976,Totokov1993} and leading to \acp{BBH} that can merge in less than the Hubble time~\citep[e.g.,][]{Bethe1998,Belczynski2002,Dominik2012,Belczynski2016b,Eldridge2016,Stevenson2017a,Giacobbo2018b}. 
However, theoretical models have shown that hardened \ac{BBH} systems merging within the Hubble time can also form through late-phase stable mass transfer~\citep{VandenHeuvel2017,Neijssel2019}. 
On the other hand, if the progenitor stars are born in a very tight orbital configuration, they may proceed through a chemically homogeneous evolution, in which rapid rotation of the stars attained through tidal interaction leads to strong mixing, replenishing the core with elements for nuclear burning and never leading to significant expansion of the progenitor stars~\citep{DeMink2016,Mandel2016b,Marchant2016}. 
Extremely low-metallicity Population III stars in binary systems have also been proposed to have formed the high-mass \ac{BBH} mergers observed~\citep{Madau2001,Kinugawa2014,Inayoshi2017}. 

On the dynamical side, following the formation of \acp{BH} from massive stars in a dense stellar environment such as a \ac{GC}, \ac{NSC}, or young open star cluster, these \acp{BH}' mass segregate to the core of the cluster due to dynamical friction and create a dense subsystem dominated by \ac{BH} interactions~\citep{Lightman1978,Sigurdsson1993a}. 
Strong gravitational encounters between \ac{BH} systems act to produce hardened binaries, typically extracting orbital energy from the more massive components of the interaction by ejecting the lighter components~\citep{McMillan1991,Hut1992,Sigurdsson1993,Miller2002,Gultekin2006,Fregeau2007} and leading to \acp{BBH} that can merge within the Hubble time~\citep[e.g.,][]{PortegiesZwart2000,OLeary2006,Downing2010,Samsing2014,Ziosi2014,Rodriguez2015a,Rodriguez2016a,Antonini2016a,Askar2017,Banerjee2017,Samsing2017c}. 
Different dynamical environments have unique predictions for the properties of merging \acp{BBH}, since stellar densities, escape velocities, and stellar mass budgets vary significantly between these environments. 

To further complicate matters, a slew of formation scenarios for the synthesis of \acp{BBH} have been proposed that do not fit cleanly into the broad categorizations of isolated binary evolution and dynamical assembly. 
A significant number of massive stars form in high-order multiples such as triples and quadruples~\citep{Sana2014,Moe2017}. 
If a \ac{BBH} is the inner binary in a hierarchical system, eccentricity can be imparted into the inner \ac{BBH} through the Lidov--Kozai mechanism~\citep{Kozai1962,Lidov1962}. 
This process will expedite the inspiral time of the binary, allowing systems to merge as \ac{GW} sources that would not typically merge within a Hubble time~\citep{Wen2003,Antonini2017a,Silsbee2017,Fragione2019,Vigna-Gomez2021}. 
Galactic nuclei are also predicted to produce \ac{BBH} mergers in a similar way, with the supermassive \ac{BH} as the outer perturber~\citep{Antonini2012}. 
Another promising environment for facilitating \ac{BBH} mergers is \acp{AGN}; \acp{BH} are predicted to get caught in resonance traps of \ac{AGN} disks, potentially proceeding through many hierarchical mergers due to the high escape velocity in the vicinity of the supermassive \ac{BH}~\citep{Mckernan2014,Bartos2017,Stone2017}. 
Stellar-mass \acp{BBH} detected by LIGO--Virgo have also been proposed to originate from the merger of central \acp{BH} in extremely low-mass ultradwarf galaxies that merge at $z \gtrsim 1$~\citep{Conselice2020}. 
Ultrawide \ac{BH} binaries and high-order systems in the galactic field that are perturbed from stellar flybys may also excite eccentricity in the \ac{BBH} system and cause it to merge within the Hubble time~\citep{Michaely2019,Michaely2020}. 
Finally, primordially formed \acp{BH} have also been proposed as sources of merging \acp{BBH} and have been suggested as candidates for dark matter~\citep{Bird2016,Sasaki2018,Clesse2020}. 

Numerous attempts have been made to leverage \ac{GW} observations for characterizing the branching fractions between these channels or constraining uncertain physical processes governing these channels~\citep[][]{Stevenson2015,Rodriguez2016,Farr2017,Farr2017a,Mandel2017,Stevenson2017,Talbot2017,Vitale2017a,Zevin2017b,Barrett2018,Taylor2018a,ArcaSedda2019a,Powell2019,Roulet2019,Wysocki2019,GWTC2_pops,Antonini2020,ArcaSedda2020,Baibhav2020,Bavera2020b,Bouffanais2020,Farmer2020,Fishbach2020b,Hall2020,Kimball2020,Kimball2020a,Roulet2020,Safarzadeh2020,Wong2020a,Wong2021,Bhagwat2021}. 
However, due to the high complexity and dimensionality of the problem, these studies often restrict themselves to targeting a single channel or a small subset of channels. 
Though idealized model comparisons are enlightening, the most robust and unbiased constraints will come from considering \emph{many prominent \ac{BBH} formation channels} and encompassing \emph{a wide range of prescriptions for uncertain physical processes}, which can affect \ac{BBH} population properties in highly degenerate ways. 

Given the rapidly growing catalog of \acp{BBH}, we are now at the stage where such high-dimensional, multichannel model selection endeavors can be informative. 
We present a methodology for leveraging a suite of state-of-the-art \ac{BBH} formation models to perform hierarchical inference using the catalog of \ac{GW} events.  
We simultaneously consider the predicted \ac{BBH} parameter distributions from five simulated populations, ensuring the models are as self-consistent as possible, and vary two uncertain physical parameters between these channels, namely the natal spins of isolated \acp{BH} and the efficiency of \ac{CE} evolution. 
Though the models we consider do not cover the entire array of proposed formation channels, the infrastructure presented in this work can be expanded to include an arbitrary number of channels and uncertain physical prescriptions. 
We find that the current catalog of \acp{GW} strongly disfavors a single formation channel, indicating a more complex landscape of prominent channels for \ac{BBH} formation. 

In Section~\ref{sec:models}, we briefly overview the astrophysical models considered in this work, as well as the physical parameterizations we vary between these models. 
Section~\ref{sec:inference} details our hierarchical modeling procedure. 
In Section~\ref{sec:results}, we show the results of our analysis applied to the current catalog of \ac{GW} observations of \ac{BBH} mergers. 
We discuss the implications of our results and conclude in Section~\ref{sec:conclusions}. 

\section{Formation Models}\label{sec:models}

We consider five astrophysical models for \ac{BBH} mergers, each with unique predictions for mass distributions, spin distributions, and the evolution of merger rate with redshift. 
Further details about these models and assumptions regarding cosmological evolution can be found in Appendix~\ref{app:pop_models}. 

\subsection{Isolated Evolution}\label{subsec:field}

Three models are considered that fall under the broad categorization of isolated evolution in the galactic field: binaries that proceed through a late-phase \ac{CE} (\texttt{CE}), binaries that only have stable mass transfer between the star and the already formed \ac{BH} (\texttt{SMT}), and chemically homogeneous evolution (\texttt{CHE}). 

The \texttt{CE} and \texttt{SMT} channels are modeled using the \texttt{POSYDON} framework (T. Fragos et al. 2021, in preparation), which, among other functionalities, stitches together different phases of binary evolution that are modeled using rapid population synthesis \citep[\texttt{COSMIC};][]{Breivik2020} and detailed binary evolution calculations \citep[\texttt{MESA};][]{Paxton2011,Paxton2013,Paxton2015,Paxton2018,Paxton2019}. 
These models are described in detail in \citet{Bavera2020b}, and the key ingredients are summarized in Appendix~\ref{app:CESMT}. 
In this channel, the more massive star leaves the main sequence first and expands to become a supergiant star. 
At some point the star overfills its Roche lobe, typically leading to a stable mass-transfer event where the primary loses most of its hydrogen envelope before it undergoes core collapse and forms a \ac{BH}. 
During the subsequent evolution the secondary expands, leading to a second mass-transfer episode. 
This can be either stable or unstable, with the latter leading to a \ac{CE} phase that shrinks the orbit more efficiently. 
If the stripping of the secondary is successful, a \ac{BH}--Wolf--Rayet system is formed, which can undergo a tidal spin-up phase. 
In general stable mass transfer leads to wider orbits compared to a \ac{CE}, and hence, the system will avoid tidal spin-up~\citep{Bavera2020b}. 
Eventually the secondary collapses to form a \ac{BH}, and energy dissipation due to \ac{GW} radiation leads to the merger.

The \texttt{CHE} models are adapted from \cite{duBuisson2020}, who computed a large grid of simulations for binaries undergoing this evolutionary process. 
Although not discussed in the work of \citet{duBuisson2020}, the simulations include predictions for the final spins of the \acp{BH}, which arise from tidal synchronization during core hydrogen burning. 
The final spin in these systems is determined by wind mass loss, which removes angular momentum and widens the binary. 
This leads to \acp{BBH} at wider separations and lower spins with increasing metallicity~\citep{Marchant2016}.

\subsection{Dynamical Assembly}\label{subsec:clusters}

We also consider two models for \ac{BBH} mergers synthesized via dynamical assembly in dense stellar environments: formation in old, metal-poor \acp{GC} (\texttt{GC}) and formation in \acp{NSC} (\texttt{NSC}). 

The \texttt{GC} simulations are taken from a grid of $96$ $N$-body models of collisional star clusters described in \cite{Rodriguez2019}.  
The models were created using the H\'enon-style cluster Monte Carlo code \texttt{CMC}~\citep{Henon1971,Henon1971a,Joshi2000,Pattabiraman2013}.  
The $96$ models consist of four independent grids of $24$ models, each with different initial spins for \acp{BH} born from stellar collapse ($\chib = 0$, $0.1$, $0.2$, and $0.5$). 
Within each $24$-model subgrid, the clusters span a range of realistic initial masses, metallicities, and half-mass radii consistent with observations of \acp{GC} in the Milky Way and nearby galaxies. 
The cluster birth times are taken from a cosmological model for \ac{GC} formation \citep{ElBadry2019}, where we take into account the correlation between cluster metallicity and formation redshift~\citep{Rodriguez2018a,Rodriguez2018c}. 

The \texttt{NSC} models are adapted from \cite{Antonini2019a}. 
For the evolution of the clusters we need the initial cluster mass $M$, half-mass radius $r_\mathrm{h}$, and \ac{BH} masses. 
Since the formation history and evolution of \acp{NSC} is uncertain \citep{Neumayer2020}, we proceed with a number of simplifying assumptions. 
We assume that the properties of nuclear clusters today are representative of their properties at formation. 
Accordingly, $M$ and $r_\mathrm{h}$ are sampled directly from the $151$ \acp{NSC} in \citet{Georgiev2014} with well-determined properties. 
For each cluster we use \texttt{COSMIC} to generate the \ac{BH} masses from a single stellar population with metallicity of $0.01 \Zsun$, $0.1 \Zsun$, or $1 \Zsun$. 
We then evolve the clusters and their \acp{BH} using the semi-analytical approach described in \citet{Antonini2019a}. 
Finally, the \ac{BBH} merger rate, masses, component spins, and redshift evolution are obtained by assuming that the formation epoch and metallicity of nuclear clusters evolve in the same way as their galactic hosts, using \cite{Madau2017a}. 
This procedure is repeated for four values of the initial \ac{BH} spins, $\chib = 0$, $0.1$, $0.2$ and $0.5$.

\begin{figure*}[t]
\includegraphics[width=0.98\textwidth]{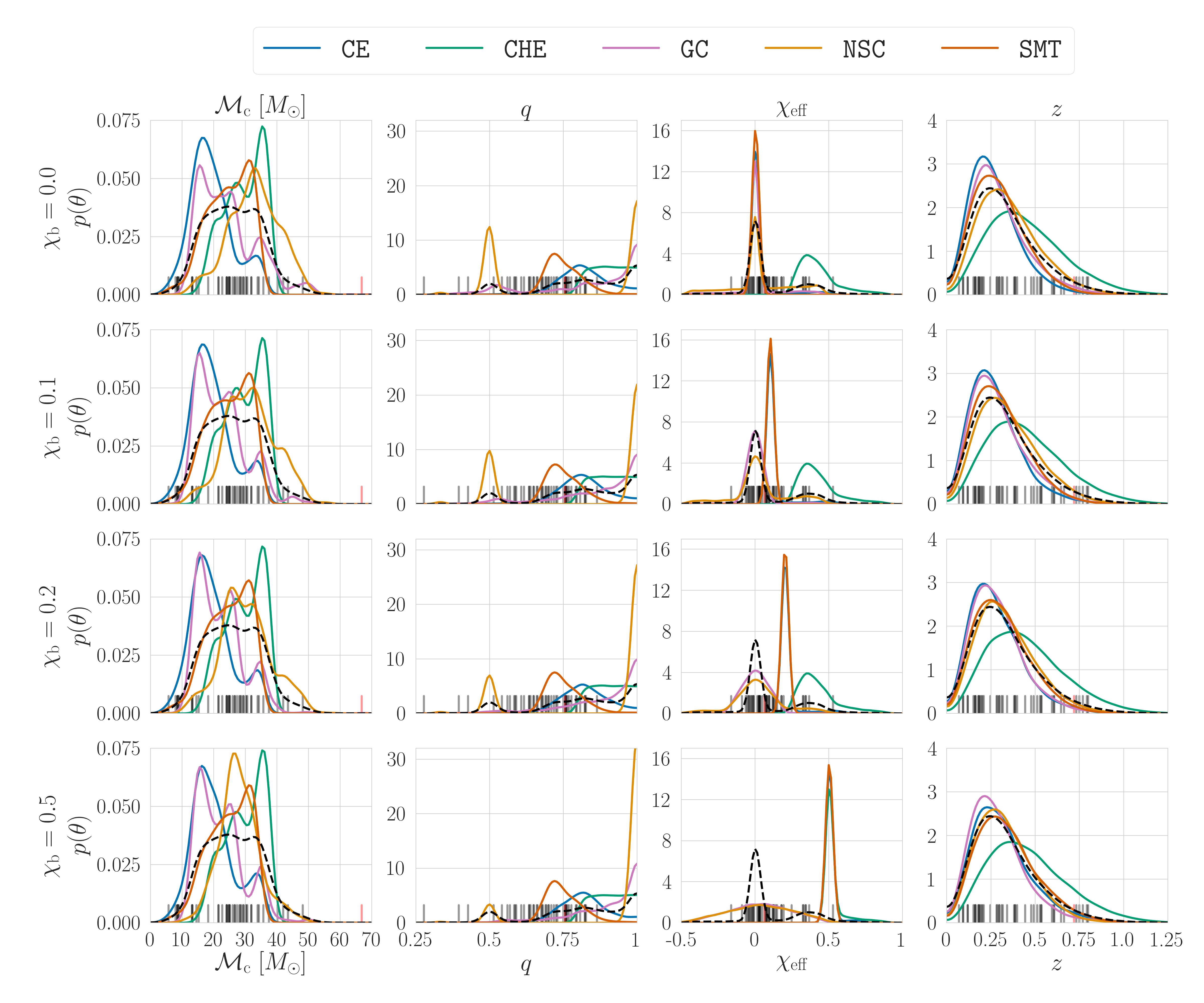}
\caption{Marginalized detection-weighted distributions of \ac{BBH} parameters for our five formation channels with varying natal spin prescriptions. 
The \ac{CE} efficiency is fixed at $\alphaCE = 1.0$ for all models in this figure. 
Black ticks mark the median value of the posterior distribution for all confident \acp{BBH} in GWTC-2; GW190521, which is excluded from our analysis, is instead marked with a red tick. 
The dashed black line shows an example mixture model synthesized for the five channels, assuming equal detectable branching fractions and a true model with the values $\chib=0.0$ and $\alphaCE=1.0$. 
}
\label{fig:pop_models}
\end{figure*}

\subsection{Physical Prescriptions}\label{subsec:prescriptions}

The physical prescription we vary across all models above is the natal spin magnitude \chib of \acp{BH} that are born in isolation or in systems where binary interactions prior to the collapse of the helium core do not cause significant spin-up. 
Variations in the natal spin magnitudes act as a proxy for the efficiency of angular momentum transport in massive stars; if angular momentum is efficiently transported from the core to the envelope, the birth spins of \acp{BH} are predicted to be low~\cite[e.g.,][]{Fuller2019b}. 
We use four models for the natal dimensionless spin magnitudes of isolated \acp{BH} in each channel: $\chib \in [0.0, 0.1, 0.2, 0.5]$. 
These discrete values for $\chib$ are chosen to match the natal spins assumed for the \ac{GC} simulations in \cite{Rodriguez2019}. 
However, this does not mean that all components of \acp{BBH} in these models are spinning at precisely these prescribed values. 

In the field channels, tidal spin-up of the pre-collapse helium core or mass transfer following the birth of the primary \ac{BH} can cause \acp{BH} to be born with or attain spin. 
For the \texttt{CE} channel, the helium core progenitor of the second-born \ac{BH} can be spun up through tidal interactions with the already born \ac{BH}. 
The degree at which the second-born \ac{BH} is spinning depends on the post-CE separation and thus on the \ac{CE} efficiency; lower \ac{CE} efficiencies will lead to tighter post-CE binaries, increasing the effect of tides and therefore increasing the natal spin of the second-born \ac{BH}~\citep{Qin2018,Zaldarriaga2018,Bavera2020b}. 
The spin of the first-born \ac{BH} can grow through stable mass transfer, though this is sensitive to assumptions regarding the maximum rate of accretion. 
Since we consider Eddington-limited accretion efficiency, the amount of spin-up that first-born \acp{BH} can acquire via accretion is minimal~\citep{Thorne1974}, and systems cannot tighten enough through this highly nonconservative mass transfer for tidal effects to be effective at spinning up the progenitor of the second-born \ac{BH}~\citep{Bavera2020b}. 
\acp{BBH} evolving through chemically homogeneous evolution are near-contact at birth, and strong tidal interaction between the stars leads to high rotations and substantial chemical mixing in both stars. 
This inhibits significant expansion of the stars, preventing efficient loss of angular momentum via accretion or loss of their envelopes. 
The natal spins of all \acp{BH} in our simulations self-consistently account for these effects, and for all three field channels considered, components spinning at $\chi < \chib$ at \ac{BBH} formation are given spins of \chib. 
Thus, unless binary interactions such as tidal effects or mass transfer are efficient at spinning up the \ac{BH} components, their spin magnitudes are assumed to be those that they would attain in isolation solely from the collapse of the stellar core. 

For the dynamical channels, the natal spins of \acp{BH} play an important role in the evolution of the \ac{BBH} subsystem in the cluster as a whole. 
In particular, the fraction of \ac{BBH} merger products retained in a cluster is highly sensitive to the spins of the \acp{BH} in the natal population~\citep{Gerosa2019a,Rodriguez2019,Kimball2020}; as spin magnitudes increase, the higher degree of asymmetry in the merger leads to larger relativistic recoil kicks due to the anisotropic emission of \acp{GW} at merger~\citep{Peres1962,Bekenstein1973,Wiseman1992,Favata2004,Baker2006a,Koppitz2007,Pollney2007,HolleyBockelmann2008,Lousto2010,Blanchet2014,Sperhake2015}. 
Thus, \ac{BBH} merger products from spinning \acp{BH}' components are more efficiently kicked out of their host environments, preventing them from proceeding in subsequent hierarchical mergers~\citep[e.g.,][]{Rodriguez2019,Banerjee2020b,Fragione2020}. 
With higher natal spins, the mass spectrum of \acp{BBH} will thus be quenched at large values by the limitations of massive-star evolution, though the retention rate and frequency of hierarchical mergers are sensitive to the mass and escape velocities of the cluster in question~\citep{Antonini2016a,Antonini2019a,Gerosa2019a,Kimball2020,Kimball2020a,Mapelli2020}. 
Suites of cluster models with varying cluster properties and metallicities are thus run for the \texttt{GC} and \texttt{NSC} channels for all spin magnitude models considered. 
Though all \acp{BH} start with the prescribed spin magnitude, higher spins in \ac{BBH} components can be attained through hierarchical mergers, which impart a spin on the newly formed \ac{BH} of $\chi \sim 0.7$ for nearly equal-mass binary mergers with non-spinning components~\citep{Pretorius2005,Gonzalez2007,Buonanno2008}. 

In addition, we consider five assumptions for the efficiency of \acp{CE}: $\alphaCE \in [0.2, 0.5, 1.0, 2.0, 5.0]$. 
Values of $\alphaCE > 1.0$ (i.e.\ efficient \ac{CE} evolution) mean that there is an energy source in addition to the orbital energy of the binary acting to remove the envelope~\citep[e.g.,][]{Ivanova2013,Nandez2016} or that some of the envelope material remains bound to the stellar core after the successful ejection of the \ac{CE}~\citep[e.g.,][]{Fragos2019}. 
These variations are assumed to affect only the \texttt{CE} channel, since the other field channels by definition do not proceed through late phases of unstable mass transfer and \acp{BBH} from dynamical channels are typically assembled after \acp{BH} form from isolated progenitor stars. 
The value of \alphaCE in the \texttt{CE} channel impacts the resultant spin distribution significantly, since tighter post-\ac{CE} binaries (lower \alphaCE) lead to more efficient tidal spin-up of the second-born \ac{BH}. 

We use a four-dimensional parameter distribution of the source-frame chirp mass \Mc, mass ratio $\q = m_2/m_1$, effective inspiral spin \chieff, and merger redshift \z in constructing the likelihoods of our population models given the \ac{GW} observations. 
Marginalized detection-weighted distributions for our population models are shown in Figure~\ref{fig:pop_models}. 
From these distributions, a variety of features from our population models can be seen. 
In the chirp mass and mass ratio distributions, varying assumptions for \chib primarily affect dynamical channels (\texttt{GC} and \texttt{NSC}); increasing \chib suppresses the high-mass bump in the chirp mass distribution and the asymmetric peak near $q \sim 0.5$, which are populated by hierarchical merger events that require lower recoil kicks and therefore lower component spins in the first-generation population. 
The asymmetric peak and high-mass bump are more pronounced in the \texttt{NSC} population than in the \texttt{GC} population since the potential well is deeper and merger products are more readily retained in the cluster. 
All formation channels show diversity in the \chieff distributions with varying \chib. 
As \chib increases, we see broader distributions for \chieff in the \texttt{GC} and \texttt{NSC} models coming from their isotropically oriented spins. 
The \texttt{CE} and \texttt{SMT} channels are strongly peaked at the prescribed value of \chib, with tails extending to higher \chieff in the \texttt{CE} channel due to systems that proceed through efficient tidal spin-up. 
The \texttt{CHE} channel typically has component spins greater than \chib and is therefore only affected in our most extreme spin scenario (\chib = 0.5). 
The redshift distributions peak at slightly larger values and broaden with increasing \chib for the \texttt{CE} and \texttt{SMT} channels since higher aligned spins spend more time in-band and are preferentially detected.

\section{Population Inference}\label{sec:inference}

Given our astrophysical models, we now establish how we place constraints on branching fractions and physical prescriptions using the current catalog of \ac{BBH} observations. 
We use posterior and prior samples from the {GWTC-1}~\citep{GWTC1} and GWTC-2~\citep{GWTC2} analyses, publicly available from the Gravitational Wave Open Science Center~\citep{GWOSC}. 
For GWTC-1 and GWTC-2, we use the \texttt{Combined} and \texttt{PublicationSamples} posterior samples, respectively, which combine posterior samples from different waveform approximants to marginalize over uncertainties in waveform modeling~\citep{GW150914_properties}. 
The choice of prior on the event parameters is irrelevant since they are divided out during the inference. 
The detection probabilities for each sample in our populations are calculated assuming a detector network consisting of LIGO Hanford, LIGO Livingston, and Virgo operating at \texttt{midhighlatelow} sensitivity~\citep{LVC_ObservingScenarios} and assuming a network \ac{SNR} threshold of $\rho_\mathrm{thresh} = 10$ (see Appendix~\ref{app:detection_probabilities}); these detection probabilities are used to construct the detection-weighted distributions in Figure~\ref{fig:pop_models}. 
We only consider high-confidence \ac{GW} events that are definite mergers of two \acp{BH}, thus excluding GW170817~\citep{GW170817}, GW190425~\citep{GW190425}, GW190426~\citep{GWTC2}, and GW190814~\citep{GW190814}. 
We also exclude GW190521~\citep{GW190521} from our analysis as it has vanishing support across our models and picks up on minute fluctuations in the \acp{KDE} for certain population models. 
This event is either not explainable by our set of channels or requires updated physical prescriptions for our set of channels. 
For each event, we randomly draw $10^2$ samples from its posterior distribution to evaluate in the population model \acp{KDE}, which are parameterized by \chib, \alphaCE, and the formation channel. 
We provide additional details for our \ac{KDE} generation in Appendix~\ref{app:kdes}. 

We perform hierarchical modeling to place constraints on the parameters influencing our population models using a methodology similar to that of \cite{Zevin2017b}. 
Since we are only interested in the shape of the populations and not in the merger rate, we implicitly marginalize over the expected number of detections~\citep[e.g.,][]{Fishbach2018}. 
The hyperparameters we wish to infer are the underlying branching fractions, ${\vec{\beta} = [\beta_{\texttt{CE}}, \beta_{\texttt{CHE}}, \beta_{\texttt{GC}}, \beta_{\texttt{NSC}}, \beta_{\texttt{SMT}}]}$, and the physical prescriptions assumed in each model, $\vec{\lambda} = [\chib, \alphaCE]$. 
We assume an uninformative prior across $\vec{\beta}$, given by a Dirichlet distribution with equal concentration parameters and dimensions equal to the number of formation channels, and impose the constraints $(0 \leq \beta_i \leq 1) \, \forall\, i$ and $\sum_i \beta_i = 1$. 
Alternatively, this prior could be proportional to the predicted local merger rates for these channels; however, given the large uncertainties in the predicted merger rates we choose an uninformative prior for this work.
We assume a uniform prior across the physical prescription parameters $\vec{\lambda}$.\footnote{In practice, we use a continuous dummy parameter to evaluate the discrete model space --- see Appendix \ref{app:inference}.} 
Since the \chib and \alphaCE models are not mixed across formation channels (i.e.\ the \texttt{CE} channel cannot have $\chib=0.0$, while the \texttt{GC} channel has $\chib=0.1$), this results in $N_{\rm channel}+1$ hyperparameters in our modeling: two physical prescriptions and ($N_{\rm channel}-1$) branching fractions, since one branching fraction is inferred given the constraint that the branching fractions sum to unity. 

Given our model hyperparameters $\vec{\Lambda} = [\vec{\lambda}, \vec{\beta}]$ and the posterior samples of our event parameters $\vec{\theta} = [\Mc, \q, \chieff, \z]$, our hyperlikelihood $p(\vec{\theta} | \vec{\Lambda})$ is given by a mixture model of channels: 
\begin{equation}\label{eq:posterior}
    p(\vec{\theta} | \vec{\Lambda}) = \sum_j \beta_j p(\vec{\theta} | \mu_j^{\chi, \alpha}), 
\end{equation}
where $\mu_j^{\chi, \alpha}$ is the (underlying) population model associated with $\beta_j$, parameterized by a particular natal spin magnitude and \ac{CE} efficiency. 
Using the discrete posterior samples for each event, the likelihood of the observed \ac{GW} data $\mathbf{x} = \{\vec{x}_{i}\}_{i}^{N_{\mathrm{obs}}}$ given our model hyperparameters is 
\begin{equation}\label{eq:hyperlikelihood}
    p(\mathbf{x} | \vec{\Lambda}) \propto 
    \prod_{i=1}^{N_\mathrm{obs}}
    \frac{1}{S_i \tilde{\xi}^{\chi, \alpha}}
    \sum_{j} \beta_j
    \sum_{k=1}^{S_i} \frac{p(\vec{\theta}_i^k | \mu_j^{\chi, \alpha})}{\pi(\vec{\theta}_i^k)}, 
\end{equation}
where $N_\mathrm{obs}$ is the number of \ac{GW} events, $S_i$ denotes the posterior samples used for event $i$, $\pi(\vec{\theta}_k)$ is the prior weight for each posterior sample in the \ac{LVC} analysis, and $\tilde{\xi}^{\chi, \alpha} \equiv \sum_j \beta_j \int  P_\mathrm{det}(\vec{\theta})p(\vec{\theta}|\mu_j^{\chi, \alpha}) \,\mathrm{d}\vec{\theta}$ is the detection efficiency of each hypermodel~\citep[e.g.,][]{Mandel2019,Vitale2020a}. 
The hyperposterior is thus given by
\begin{equation}
    p(\vec{\Lambda} | \mathbf{x}) = p(\mathbf{x} | \vec{\Lambda}) \pi(\vec{\Lambda}),
\end{equation}
where $\pi(\vec{\Lambda})$ is the prior on our hyperparameters. 

In practice, the likelihood of Eq.~\eqref{eq:hyperlikelihood} is evaluated by first moving in the (discrete) physical prescription parameter $\vec{\lambda}$ space, then moving in the (continuous) branching fraction $\vec{\beta}$ space, and evaluating if the jump proposal is accepted. 
Thus, Eq.~\eqref{eq:hyperlikelihood} consists of evaluations from a mixture model of the underlying population \acp{KDE} for the given values of \chib, \alphaCE, and $\vec{\beta}$ at a particular step in the sampler. 
We use the ensemble sampler from \texttt{emcee}~\citep{Foreman-Mackey2013} to sample this distribution, and perform $10^2$ realizations of this inference with different random samplings from the event posteriors, creating a combined posterior distribution across these realizations. 
Further details can be found in Appendix~\ref{app:inference}, and results from a mock injection study using this methodology can be found in Appendix~\ref{app:mock_obs}.

\section{Application to GWTC-2}\label{sec:results}

\subsection{Two-channel example}\label{subsec:twochannel}

\begin{figure}[t]
\includegraphics[width=0.48\textwidth]{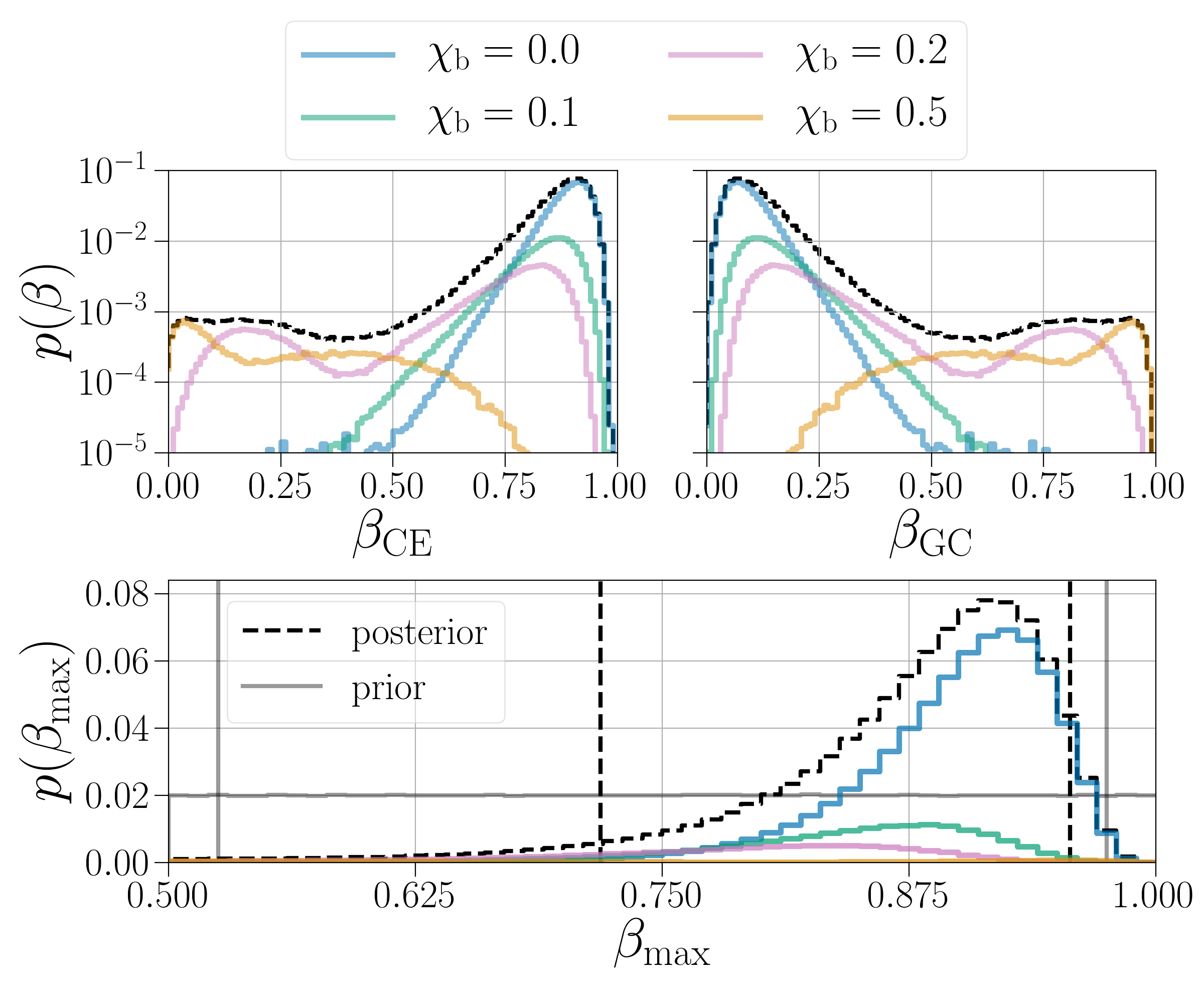}
\caption{\textit{Top row}: Branching fractions between \texttt{CE} and \texttt{GC} channels, under the assumption that only these two channels contribute to the \ac{BBH} catalog. 
Black dashed lines show the posterior on the detected branching fractions $\beta$, marginalized over the \chib and \alphaCE models. 
Colored lines show the contributions to the full $\beta$ posterior from various \chib models. 
\textit{Bottom row}: Posterior distribution on $\beta_\mathrm{max} = \max(\vec{\beta})$. 
The gray line shows the prior distribution for $\beta_\mathrm{max}$; vertical lines mark the symmetric 90\% credible interval for the fully marginalized posterior and prior distributions. 
}
\label{fig:beta_CEGC}
\end{figure}

We first consider a simplified picture to build intuition for the full analysis, assuming that the \ac{BBH} population comes from only the \texttt{CE} and \texttt{GC} channels. 
The posterior distributions for the underlying branching fractions $\beta$ are shown in Figure~\ref{fig:beta_CEGC}, with colored lines showing the contribution of various \chib models to the full posteriors for $\vec{\beta}$. 
In this simplified case, we already see some notable features. 
The Bayes factors \BF between models are given by the number of samples in one physical prescription model compared to another (i.e.\ the relative area under the colored curves in the top panels of Figure~\ref{fig:beta_CEGC}). 
We find a preference for our smallest natal spin model ($\chib = 0.0$) relative to the higher \chib models considered. 
Compared to the highest natal spin model ($\chib=0.5$), the $\chib=0.0$ model is preferred by a Bayes factor of $\BF_{\chib\,=\,0.5}^{\chib\,=\,0.0} \simeq \BFchifiveCEGC$. 
However, once natal spin magnitudes are decreased to lower values, the preference becomes more marginal, $\BF_{\chib\,=\,0.2}^{\chib\,=\,0.0} \simeq \BFchitwoCEGC$ and $\BF_{\chib\,=\,0.1 }^{\chib\,=\,0.0} \simeq \BFchioneCEGC$, respectively. 
This is consistent with the population analysis associated with {GWTC-2}, which pushes to low component spins for the \ac{BBH} population~\citep{GWTC2_pops}.  

When marginalizing over all values of \chib and \alphaCE, the median and symmetric $90\%$ credible interval of the posterior distributions for $\beta_{\texttt{CE}}$ and $\beta_{\texttt{GC}}$ are $\BetaCETwochannel$ and $\BetaGCTwochannel$, respectively. 
The branching fractions are sensitive to the value of \chib; we find a significant increase (decrease) in $\beta_{\texttt{GC}}$ ($\beta_{\texttt{CE}}$) for models with $\chib \geq 0.1$. 
If we take our most extreme natal spin model, $\chib = 0.5$, the inferred branching fractions reverse: $\beta_{\texttt{CE}} = \BetaCETwochannelChiFive$ and $\beta_{\texttt{GC}} = \BetaGCTwochannelChiFive$.
However, this model is strongly disfavored by the data. 
In the bottom panel of Figure~\ref{fig:beta_CEGC}, we also gauge the preference for a mixture of channels in the underlying population by evaluating the posterior distribution for $\beta_\mathrm{max}$, defined as the largest value of $\beta$ across all channels. 
We find $\beta_\mathrm{max} \lesssim \MaxBetaNinetyCEGC$ ($\MaxBetaNinetyNineCEGC$) at the 90\% (99\%) credible level when all spin models are considered. 
For this simplified case, if natal spins for \acp{BH} born in isolation are low, which is favored by the data, the \texttt{CE} channel dominates the underlying \ac{BBH} population, though some contribution from the \texttt{GC} channel is still necessary. 
These results are for the underlying population of \acp{BBH}; we discuss the conversion to the detectable population in the following section.

\subsection{Five-channel analysis}\label{subsec:fivechannel}

\begin{figure*}[t]
\includegraphics[width=0.98\textwidth]{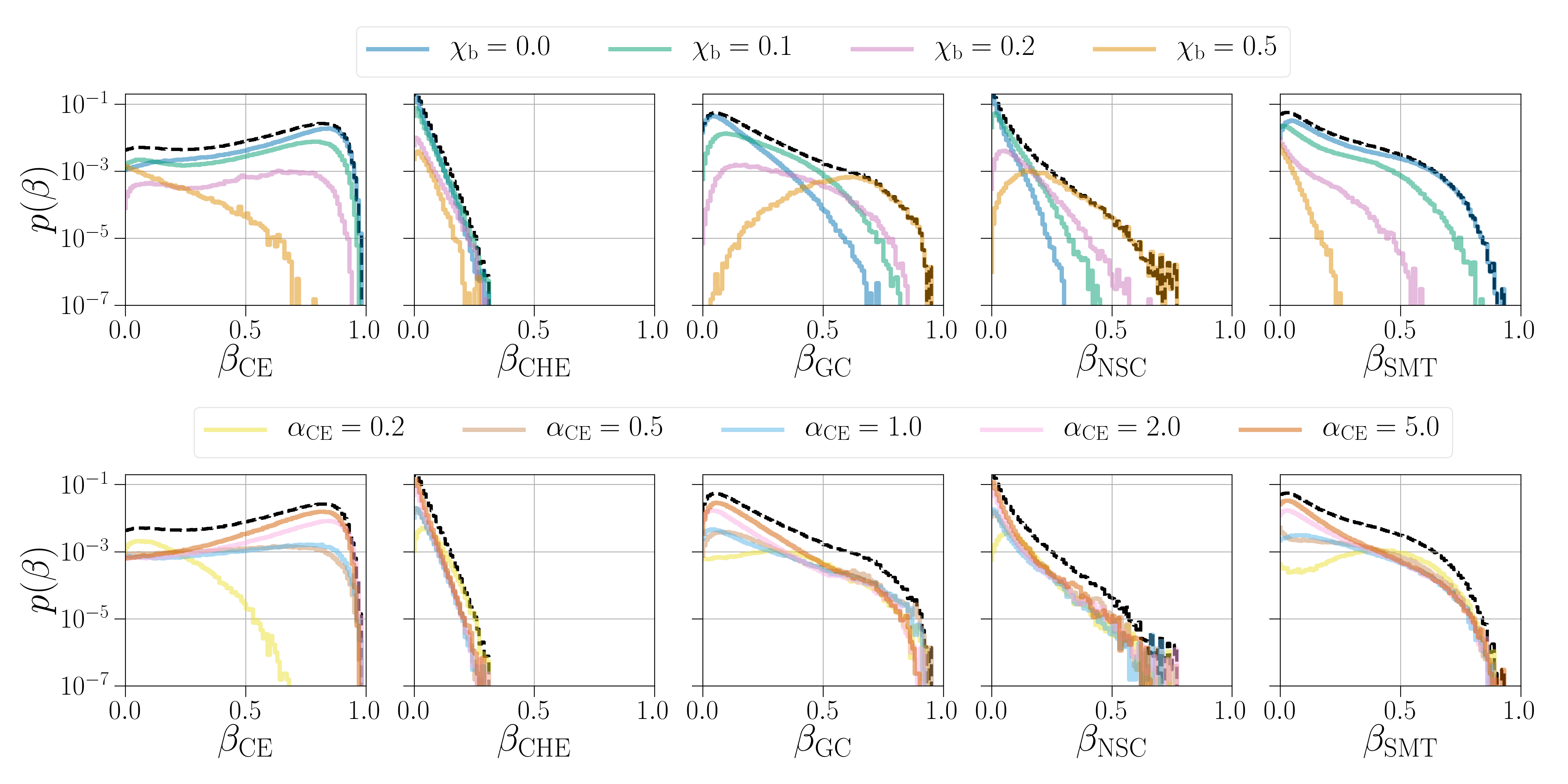}
\caption{Branching fractions for all five channels inferred using the GWTC-2 \acp{BBH}. 
Colored lines show the contributions from various \chib models marginalized over $\alphaCE$ models (\textit{top row}), and various $\alphaCE$ models marginalized over \chib models (\textit{bottom row}). 
Black dashed lines show the full posterior on the branching fractions, marginalized over both \chib and \alphaCE models. 
}
\label{fig:beta_posteriors}
\end{figure*}

\begin{figure*}[t]
\includegraphics[width=0.98\textwidth]{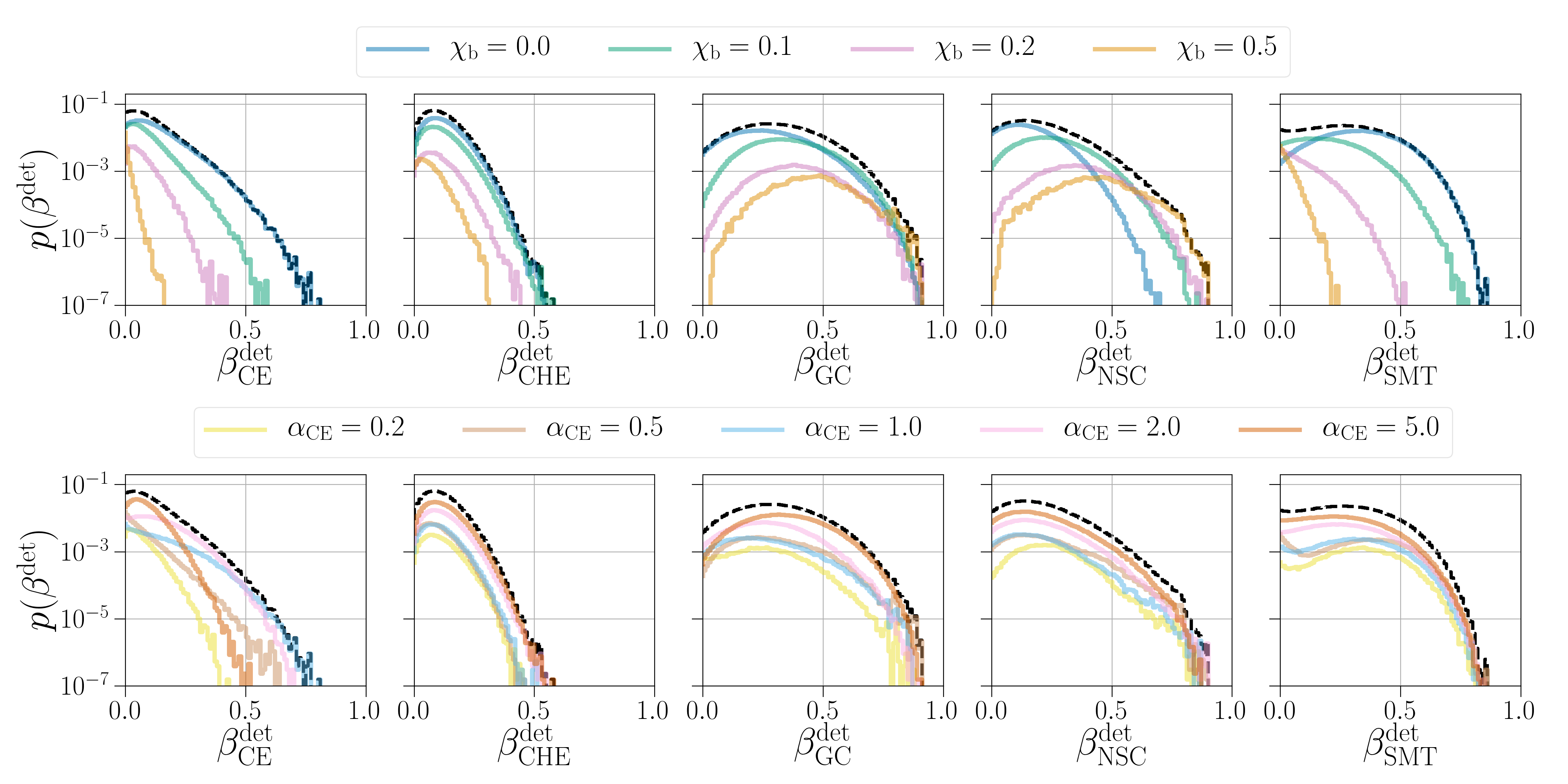}
\caption{Same as Figure~\ref{fig:beta_posteriors}, but with detectable branching fractions instead of branching fractions for the underlying population. 
}
\label{fig:beta_posteriors_detectable}
\end{figure*}

\begin{figure}[t]
\includegraphics[width=0.45\textwidth]{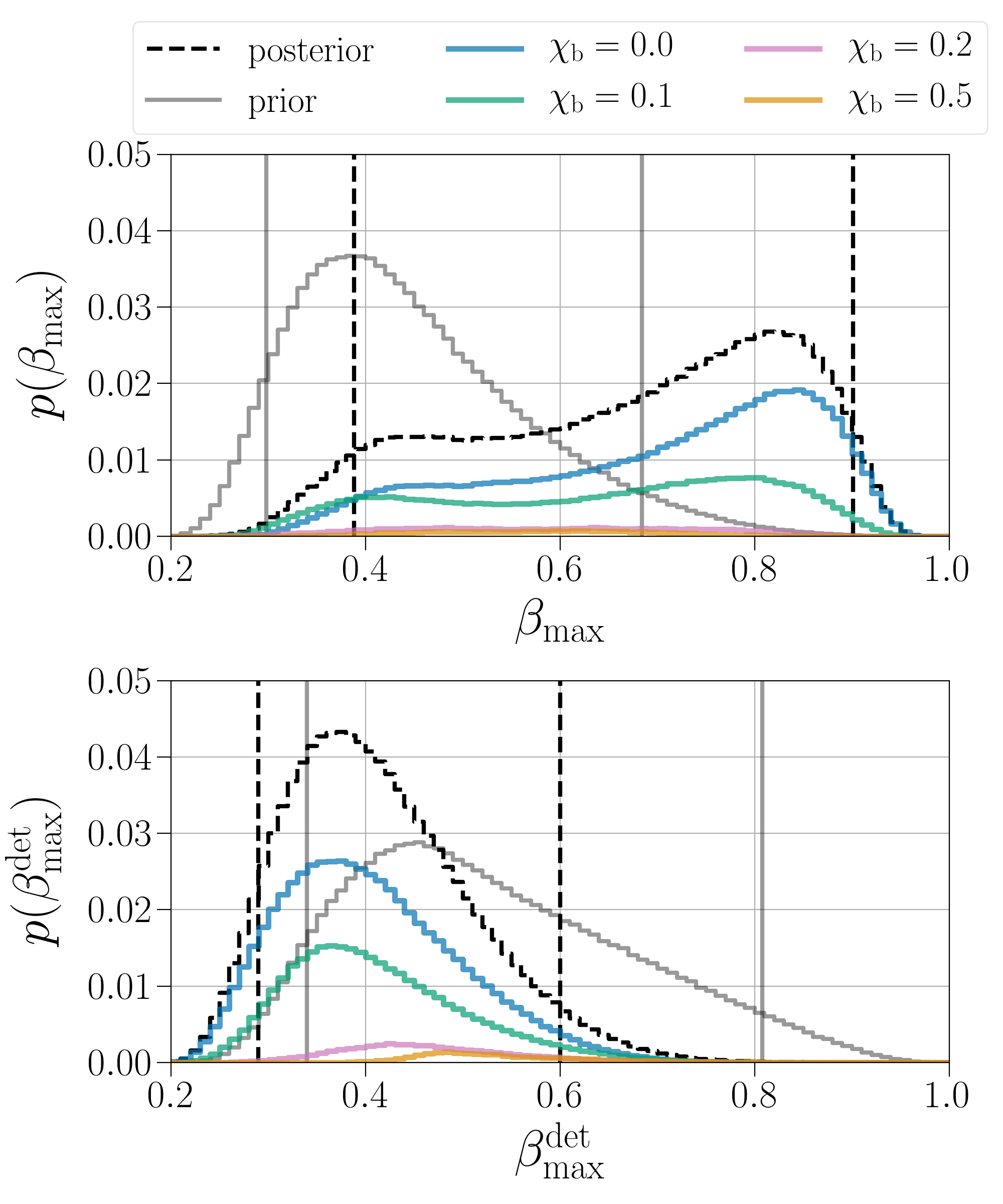}
\caption{Posterior distribution for the maximum branching fractions of the underlying population $\beta_\mathrm{max}$ (\textit{top row}) and detectable population $\beta_\mathrm{max}^\mathrm{det}$ (\textit{bottom row}) when considering all five formation channel models. 
The black dashed line shows the fully marginalized posterior distribution, with colored lines showing the contribution to the posterior from the different \chib models marginalized over all \alphaCE models. 
We also show the prior distributions for $\beta_\mathrm{max}$ and $\beta_\mathrm{max}^\mathrm{det}$ with gray lines; vertical black dashed lines and gray lines mark the symmetric 90\% credible interval for the fully marginalized posterior and prior, respectively. 
There is a clear difference between the branching ratios in the detectable population and the underlying population. 
Since higher-mass binaries are more likely to be detected, the subpopulations that contribute to these channels are enhanced, making the detectable population much more diverse. 
}
\label{fig:beta_max}
\end{figure}

We now perform the same analysis but using all of our five formation models. 
In this case, we also consider the branching fractions for the detectable population $\vec{\beta}^\mathrm{det}$ which encodes the breakdown of formation channels to the population detected by LIGO--Virgo. 
To transform to such \textit{detectable} branching fractions, we rescale the recovered underlying branching fractions by the detection efficiency of each population model, $\xi^{\chi, \alpha}_j = \int  P_\mathrm{det}(\vec{\theta})p(\vec{\theta}|\mu_j^{\chi, \alpha})\,\mathrm{d}\vec{\theta}$: 
\begin{equation}\label{eq:detectable_conversion}
    \vec{\beta}^\mathrm{det} = \left[ \frac{\vec{\beta} \odot \vec{\xi}}{\tilde{\xi}}\right]_{\chi, \alpha},
\end{equation}
where $\vec{\xi}$ is a vector of detection efficiencies for all formation models $j$ given a particular submodel $\chi, \alpha$ and $\vec{\beta} \odot \vec{\xi}$ is the element-wise product of $\vec{\beta}$ and $\vec{\xi}$ for samples in the submodel $\chi, \alpha$. 
The posterior distributions for the underlying and detectable branching fractions are shown in Figures~\ref{fig:beta_posteriors} and \ref{fig:beta_posteriors_detectable}, respectively, with the top row breaking down the contribution from the various \chib models and the bottom row breaking down the contribution from the various \alphaCE models. 

Even when considering all five channels, the underlying populations may be dominated by the \texttt{CE} channel, though there is significant support for lower values of $\beta_{\texttt{CE}}$ and a non-negligible contribution from the other channels (see Figure~\ref{fig:beta_posteriors}). 
The median and 90\% credible interval for the underlying branching fraction posteriors $\vec{\beta} = [\beta_{\texttt{CE}}, \beta_{\texttt{CHE}}, \beta_{\texttt{GC}}, \beta_{\texttt{NSC}}, \beta_{\texttt{SMT}}]$ are [$\BetaCE$, $\BetaCHE$, $\BetaGC$, $\BetaNSC$, $\BetaSMT$]. 
These correspond to relative measurement uncertainties on the underlying branching fractions of $\sim\,\PrecentLevelConstraintsLow$--$\PrecentLevelConstraintsHigh$ ($90\%$ credibility). 
Though most of the distributions are broad, the \texttt{CE} channel has the least support at $\beta = 0$; at least $\BetaCEOnePercent$ of the underlying population is from the \texttt{CE} channel at $99\%$ credibility. 

As expected, we also find notably different inferred branching fractions and model preferences between the two-channel case and the five-channel case. 
For example, as compared to that in the two-channel analysis, in the five-channel analysis the median branching fraction for the \texttt{CE} channel decreases by $\PrecentDifferenceCE$ while the median branching fraction for the \texttt{GC} channel increases by $\PrecentDifferenceGC$, and the $\chib=0.1$ model is increasingly favored by a factor of $\BF_{\chib\,=\,0.1}^\mathrm{5-channel} / \BF_{\chib\,=\,0.1}^\mathrm{2-channel} \simeq \ChiOneBFCompare$. 

In certain formation channels, we see the branching fractions converge to different values when different physical prescriptions are assumed. 
The branching fractions for dynamical (isolated) channels push to larger (smaller) values with increasing \chib; moving from $\chib=0.0$ to $\chib=0.5$ leads to an increase in the median recovered branching fraction of $\BetaIncreaseGC$ for the \texttt{GC} channel. 
This is due to the effective inspiral spin distribution for the \acp{BBH} in the \ac{LVC} catalog, which is near-symmetric about zero although slightly skewed toward positive values~\citep{GWTC2_pops}. 
As natal spins increase, the isotropic spin orientations in dynamical environments lead to broader and symmetric effective inspiral spin distributions, whereas the relatively aligned spins from isolated evolution channels lead to a strong peak in the effective inspiral spin distributions at positive values (see Figure~\ref{fig:pop_models}). 
Thus, under the assumption of nonzero natal spins, the inferred relative branching fraction between channels will change, increasing the relative contribution from dynamical formation channels. 
Systematic shifts in the inferred branching fractions are also apparent for variations in the \ac{CE} efficiency; for the \texttt{CE} channel, changing \alphaCE by an order of magnitude from $0.2$ to $2.0$ increases the median $\beta_{\texttt{CE}}$ by $\BetaIncreaseCE$. 

The posterior distributions on detectable branching fractions are shown in Figure~\ref{fig:beta_posteriors_detectable}. 
The median and $90\%$ credible interval for the detectable branching fraction posteriors $\vec{\beta}^\mathrm{det} = [\beta_{\texttt{CE}}^\mathrm{det}, \beta_{\texttt{CHE}}^\mathrm{det}, \beta_{\texttt{GC}}^\mathrm{det}, \beta_{\texttt{NSC}}^\mathrm{det}, \beta_{\texttt{SMT}}^\mathrm{det}]$ are [$\BetaDetCE$, $\BetaDetCHE$, $\BetaDetGC$, $\BetaDetNSC$, $\BetaDetSMT$]. 
The detectable branching fractions for all channels other than \texttt{CE} increase relative to the underlying branching fractions, since the \texttt{CE} channel typically produces \acp{BBH} with lower masses. 
This is due to the mass spectrum of these channels pushing to larger values, particularly for the favored $\chib = 0$ model, which leads to an increase in their detection efficiency. 
Given our astrophysical models, the \texttt{GC} channel contributes to the bulk of the detected population, making up $> \BetaDetectableGCNinetyNine$ of the observed population at $99\%$ credibility. 
The most significant increases when converting to detectable branching fractions are in the channels whose mass spectra push to the largest values; given our set of formation models, the median detectable branching fraction for the \texttt{NSC} channel is almost an order of magnitude larger than the median underlying branching fraction. 

Once again, the detectable branching fraction for isolated channels rises with decreasing natal spin magnitude. 
For example, the median value for $\beta_{\texttt{SMT}}^\mathrm{det}$ increases by $\BetaDetectableSMTIncrease$ when considering the $\chib = 0.0$ model relative to the fully marginalized models. 

To further gauge whether a single channel or multiple channels are favored by the \ac{BBH} population, we again show the posterior distribution on $\beta_\mathrm{max}$ in Figure~\ref{fig:beta_max}, now with all five formation channels included. 
In this higher-dimensional case, the prior on $\beta_\mathrm{max}$ has a more complicated morphology; the prior volume near $\beta  = 1$ for any one channel drops precipitously. 
However, we still see that the posterior on $\beta_\mathrm{max}$ significantly deviates from the prior, pushing to larger values of $\beta_\mathrm{max}$ and favoring one channel dominating the underlying population. 
For the underlying branching fractions, we find that $\beta_\mathrm{max}$ is constrained to be below $\PosteriorNinetyBmax$ ($\PosteriorNinetyNineBmax$) at the $90\%$ ($99\%$) credible level, compared to $\PriorNinetyBmax$ ($\PriorNinetyNineBmax$) for the prior. 
Conversely, we find a mixture of channels contributing to the detected population to be strongly preferred. 
For the detectable population, $\beta_\mathrm{max}^\mathrm{det} < \PosteriorNinetyBmaxDetectable$ ($\PosteriorNinetyNineBmaxDetectable$) at $90\%$ ($99\%$) credibility compared to $\PriorNinetyBmaxDetectable$ ($\PriorNinetyNineBmaxDetectable$) for the prior. 

Another metric we can consider is the number of channels that dominate the branching fractions. 
We gauge this by examining the posterior distribution on the number of branching fractions that are simultaneously above a threshold value. 
Setting the threshold to $\beta = 0.1$ and marginalizing over all \chib and \alphaCE models, we find that $\simeq \NOneBetaThreshold$ of the posterior for the underlying branching fractions supports a single model contributing to more than $10\%$ of the underlying population, and $> \NLeqThreeBetaThreshold$ of the posterior supports a significant contribution from three or fewer channels. 
Though our \texttt{CE} model is favored, this indicates that there may still be an appreciable contribution from a couple of other formation channels to the underlying \ac{BBH} population. 
This picture changes drastically for the detectable population. 
We find that there is a $> \NGtrOneThresholdDetectable$ probability that more than one channel contributes to at least $10\%$ of the detected population, with the bulk of the posterior support ($\simeq \NThreeFourThresholdDetectable$) suggesting that three to four channels significantly contribute. 
Thus, given our astrophysical models, the detected catalog of \acp{BBH} comes from a diverse array of formation scenarios. 

The Bayes factors between the physical prescriptions $\vec{\lambda} = [\chib, \alphaCE]$ are given in Table~\ref{tab:table}, analogous to the fully integrated colored curves in Figure~\ref{fig:beta_posteriors}. 
As with the two-channel case, we find moderate to strong preference for low natal spins of $\chib \lesssim 0.1$ relative to larger natal spins, in agreement with other work investigating natal spin distributions using the catalog of \ac{BBH} events ~\citep{Farr2017a,GWTC2_pops,Kimball2020a,Miller2020}. 
Spins of $\chib \lesssim 0.1$ are favored relative to models with $\chib \gtrsim 0.2$ by a Bayes factor of $\mathcal{B} \simeq \BFchilowchihigh$. 
The marginal preference for the no-spin $\chib = 0.0$ model compared to the low-spin $\chib = 0.1$ model ($\mathcal{B}^{\chib = 0.0}_{\chib = 0.1} \simeq \BFchizerochione$) indicates no strong discriminating power between the two. 

We also marginally prefer high \ac{CE} efficiencies of $\alphaCE \simeq 5.0$, which have Bayes factors of $\mathcal{B} \simeq \BFalphafive$ relative to the $\alphaCE = 1.0$ model. 
Values of $\alphaCE=1.0$ and $\alphaCE=0.5$ show near-equal preference, and highly inefficient \acp{CE} with $\alphaCE = 0.2$ are disfavored relative to the most highly favored model ($\alphaCE = 5.0$) by a Bayes factor of $\gtrsim\,\BFalphalowest$. 
Though we use the \texttt{CE} and \texttt{SMT} models from \cite{Bavera2020b} in this work, we find the opposite results in terms of the inferred \ac{CE} efficiency. 
In \cite{Bavera2020b}, low \ac{CE} efficiencies preferentially form more massive \acp{BBH}. 
Since only the \texttt{CE} and \texttt{SMT} channels are considered in \cite{Bavera2020b}, the preference for low \ac{CE} efficiencies comes from the necessity to produce these more massive systems to match the properties of the events in {GWTC-2}, whereas in this work such systems can be explained by alternative formation channels. 
We tested this hypothesis by considering two-channel inference that included only the \texttt{CE} and \texttt{SMT} channels, and also found a strong preference for \ac{CE} efficiencies of $\alphaCE \simeq 0.5$, with no samples in the efficient ($\alphaCE > 1.0$) \ac{CE} models.

\begin{table}[t]
\begin{center}
\caption{Bayes factors $\log_{10}(\mathcal{B})$ across \chib models (columns) and \alphaCE models (rows). 
Bayes factors are normalized against $\alphaCE=1.0$ and $\chib=0.0$ in the general case, against $\chib=0.0$ when marginalizing over \alphaCE, and against $\alphaCE=1.0$ when marginalizing over \chib. 
The bottom row provides the Bayes factors for \chib models marginalized over all \alphaCE models, and the rightmost column provides the Bayes factors for \alphaCE models marginalized over all \chib models. 
\label{tab:table}}
\setlength{\tabcolsep}{6pt}
\vspace{-10pt}
\begin{tabular}{c c c c c c c} 
& & \multicolumn{4}{c}{$\chi_{\rm b}$} & \\
& \multicolumn{1}{l|}{} & $0.0$ & $0.1$ & $0.2$ & \multicolumn{1}{l|}{$0.5$} & \\
\cline{2-7}
\multirow{5}{*}{$\alpha_{\rm CE}$} & \multicolumn{1}{l|}{$0.2$} & $-0.63$ & $-0.56$ & $-1.24$ & \multicolumn{1}{l|}{$-1.71$} & $-0.35$ \\
& \multicolumn{1}{l|}{$0.5$} & $-0.06$ & $-0.58$ & $-0.96$ & \multicolumn{1}{l|}{$-1.11$} & $0.00$ \\
& \multicolumn{1}{l|}{$1.0$} & $\equiv0$ & $-0.77$ & $-1.02$ & \multicolumn{1}{l|}{$-1.29$} & $\equiv0$ \\
& \multicolumn{1}{l|}{$2.0$} & $0.34$ & $0.05$ & $-1.19$ & \multicolumn{1}{l|}{$-1.15$} & $0.42$ \\
& \multicolumn{1}{l|}{$5.0$} & $0.56$ & $0.39$ & $-0.54$ & \multicolumn{1}{l|}{$-0.87$} & $0.70$ \\
\cline{2-7}
& \multicolumn{1}{l|}{$$} & $\equiv0$ & $-0.27$ & $-1.11$ & \multicolumn{1}{l|}{$-1.35$} & $$ 
\end{tabular}
\end{center}
\end{table}

\section{Discussion and Conclusions}\label{sec:conclusions}

We analyze the recently bolstered catalog of \ac{BBH} mergers using a suite of state-of-the-art models for astrophysical formation channels of \acp{BBH}. 
Our main findings are as follows: 
\begin{itemize}
    \item Though the \ac{CE} channel dominates the underlying \ac{BBH} population in our models, a contribution from various formation channels is preferred over one channel dominating the detected population of \ac{BBH} mergers. 
    From the formation channels considered in this work, we find that no single channel contributes to more than $\PosteriorNinetyNineBmaxDetectablePercentApprox$ of the detectable \ac{BBH} population at 99\% credibility, and the probability that three to four channels each contribute to more than 10\% of the detected \ac{BBH} population is $\NThreeFourThresholdDetectable$. 
    \item Small natal spins ($\chib \lesssim 0.1$) for \acp{BH} born in isolation or without significant tidal influence from a binary partner are favored over larger natal spins ($\chib \gtrsim 0.2$) by a Bayes factor of $\simeq \BFchilowchihigh$, indicating efficient angular momentum transport in massive stars. 
    \item The \ac{CE} efficiency, which scales roughly linearly with the post-\ac{CE} separation, shows marginal preference for larger values ($\alphaCE \simeq 5.0$) relative to $\alphaCE \approx 1.0$ by a Bayes factor of $\approx\,\BFalphafive$, and stronger preference relative to highly inefficient \acp{CE} ($\alphaCE \simeq 0.2$) by a Bayes factor $\gtrsim\,\BFalphalowest$. 
    This preference for efficient \ac{CE} ejection may indicate that other energy sources are at play when ejecting \acp{CE} rather than solely the orbital energy of the binary~\citep{Ivanova2013}. 
    \item When incorrect physical prescriptions are assumed or formation channels contributing to the \ac{BBH} population are not considered, estimates for the values of branching fractions and variables in physical parameterizations can be significantly biased.
\end{itemize}

Numerous studies have investigated how populations of compact-object mergers can help inform uncertainties in binary stellar evolution and compact-object formation. 
This is typically done with either phenomenological models or predictions from population models, though in the case of the latter, usually under the assumption that a single population model is exclusively contributing to the entire population. 
However, several studies have considered the relative contribution and population properties expected from dynamical channels versus isolated binary evolution~\citep[e.g.,][]{Rodriguez2016,Farr2017a,Stevenson2017,Vitale2017a,Zevin2017b,Bouffanais2019,ArcaSedda2020,Safarzadeh2020b,Santoliquido2020,Wong2020a}. 
For example, when considering \ac{BBH} mergers formed in isolation or dynamically in young stellar clusters, \cite{Bouffanais2019} found that branching fractions can be constrained to the $\sim\,10\%$ level with $\mathcal{O}(100)$ detections. 
With only $\sim\,50$ observations in GWTC-2 and the inclusion of additional formation channels, we see broader constraints on the precise values of branching fractions; investigating the convergence on branching fraction estimates when considering more channels will be investigated in future work. 
\cite{Wong2020a} also analyzed the \ac{BBH} population from GWTC-2 using models for the \ac{GC} and \ac{CE} channels.  
We find opposite results for the underlying branching fractions when we consider the simplified picture of only the \texttt{CE} and \texttt{GC} channels (see Figure~\ref{fig:beta_CEGC}); we find the majority of the underlying population ($\sim 90\%$) comes from the \ac{CE} channel, whereas \cite{Wong2020a} found $\sim 80\%$ of the underlying population comes from the \ac{GC} channel. 
However, our detectable distributions are in better agreement with the branching fractions presented in \cite{Wong2020a}; incorporating detectability increases the detectable branching fraction of \acp{GC} in the two-channel example to $\sim 70\%$. 
We also find a slight preference for efficient \acp{CE} and a stronger preference against highly inefficient \acp{CE} ($\alphaCE \simeq 0.2$), whereas the constraints for \alphaCE in \cite{Wong2020a} are broad but slightly favor inefficient \acp{CE}. 
The difference in our analyses may be due to our inclusion of spin information, since the \ac{CE} efficiency has a stronger impact on the spin distributions of the \ac{CE} channel compared to the mass spectrum. 
This emphasizes the importance of considering all observational information when constraining models.

\begin{figure}[b]
\includegraphics[width=0.46\textwidth]{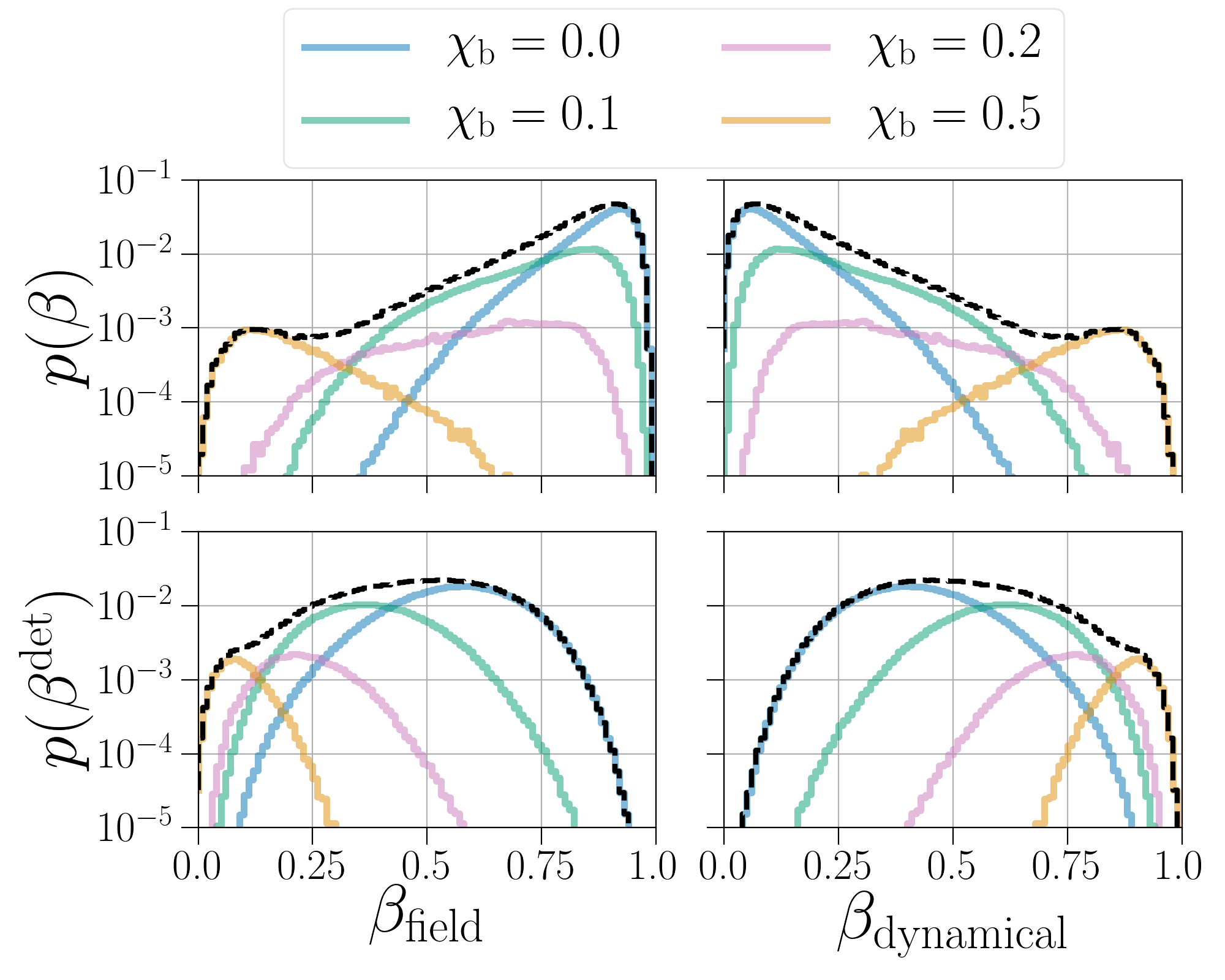}
\caption{Branching fractions recovered for the combined field and dynamical channels. 
Colored lines show the contribution to the posterior from the different \chib models marginalized over all $\alphaCE$ models. 
In the top row we show the distribution for underlying branching fractions as in Figure~\ref{fig:beta_posteriors}, and in the bottom row we show the distribution for detectable branching fractions as in Figure~\ref{fig:beta_posteriors_detectable}. 
}
\label{fig:field_dynamical}
\end{figure}

Though we consider five distinct \ac{BBH} formation models in this analysis, we can also investigate the broad two-channel categorization of formation in the galactic field and dynamical assembly in dense stellar environments. 
In Figure~\ref{fig:field_dynamical}, we combine the branching fractions for the field channels (\texttt{CE}, \texttt{CHE}, \texttt{SMT}) against the dynamical channels (\texttt{GC}, \texttt{NSC}). 
With low spins ($\chib \lesssim 0.1$), we find a strong preference for field channels comprising the majority of the underlying distribution. 
The field channels are dominated by the \texttt{CE} channel, which makes up the majority of the underlying population (see Figure~\ref{fig:beta_posteriors}). 
Marginalizing over all \chib and \alphaCE models, we find the underlying branching fractions for the field channels and dynamical channels are $\beta_\mathrm{field} = \BetaField$ and $\beta_\mathrm{dynamical} = \BetaDynamical$, respectively (90\% credibility). 
The contribution from dynamical channels increases as natal spins increase due to the behavior of the effective inspiral spin distributions, since the effective inspiral spins of the \ac{BBH} events in GWTC-2 are near-symmetric about zero~\citep{GWTC2_pops} and incompatible with a highly spinning, aligned-spin population. 
At $\chib = 0.5$, which is disfavored relative to $\chib = 0$ by a Bayes factor of $\simeq \BFchizerochifive$, the underlying branching fractions are $\beta_\mathrm{field} = \BetaFieldChiFive$ and $\beta_\mathrm{dynamical} = \BetaDynamicalChiFive$. 
In the bottom panel of Figure~\ref{fig:field_dynamical}, we again use Eq.~\eqref{eq:detectable_conversion} to convert underlying branching fractions to detectable branching fractions. 
In the detectable population, the contribution from dynamical channels is amplified. 
Marginalizing over all \chib and \alphaCE values, we find the inferred branching fractions of the detected populations to be $\beta_\mathrm{field}^\mathrm{det} = \BetaFieldDet$ and $\beta_\mathrm{dynamical}^\mathrm{det} = \BetaDynamicalDet$. 
Thus, given the formation models considered in this work, dynamical and field channels contribute similar numbers to the detected \ac{BBH} population. 

Our analysis favors low natal spins for \acp{BH} born in isolation or without significant tidal spin-up. 
Analyses using phenomenological representations for the spin magnitude distribution show a preference for low spins~\citep{Farr2017a,O2RandP,GWTC2_pops,Kimball2020a,Miller2020}, in agreement with the preference for low natal spins in this work. 
Low spins in the natal population will increase the rate of hierarchical mergers in dynamical environments~\citep{Rodriguez2019,Banerjee2020b,Fragione2020,Fragione2021,Kimball2020}, pushing the \ac{BH} mass spectrum to larger values, imparting large spin on the merger products, and accentuating the mass asymmetry of mergers in those populations. 

We find a mild preference for efficient \acp{CE} in our modeling, which strongly disfavors highly inefficient \acp{CE} with $\alphaCE \simeq 0.2$. 
Inferred values for the \ac{CE} efficiency \alphaCE have more diversity than natal spins across the literature, from population modeling~\citep[e.g.,][]{Bavera2020b,Santoliquido2021,Zevin2020b} to hydrodynamical simulations~\citep[e.g.,][]{Fragos2019} and theoretical considerations~\citep[e.g.,][]{Ivanova2013}. 
Our results, which mildly favor high \ac{CE} efficiencies, are in agreement with \citet{Santoliquido2021}, who found high \ac{CE} efficiencies are necessary to match the merger rate of binary neutron stars in their population models, and with \cite{Fragos2019}, who modeled the spiral-in phase of \ac{CE} evolution using hydrodynamic simulations and found a non-negligible fraction of the envelope remains bound to the core after the \ac{CE} is successfully ejected. 
These results for \alphaCE contrast with those from \cite{Bavera2020b}, but we find that these conflicting results arise only from the consideration of more formation channels in this work; when considering contributions from only the \texttt{CE} and \texttt{SMT} channels, we also favor low \ac{CE} efficiencies of $\alphaCE \simeq 0.5$. 
We also find similar detectable branching fractions for the \texttt{CE} and \texttt{SMT} channels, in agreement with \cite{Bavera2020b} who found the channels have comparable \ac{BBH} detection rates in the local universe. 

We have shown multiple times that failing to account for the broad array of formation channels or assuming incorrect physical prescriptions can severely bias inferences. 
For example, when only considering the \texttt{CE} and \texttt{SMT} channels in our inference, we find a preference for low \ac{CE} efficiencies compared to the preference for high \ac{CE} efficiencies when considering all five channels. 
When considering only the \texttt{CE} and \texttt{GC} channels, the recovered branching fractions differ significantly compared to when we consider all five channels (see Figures~\ref{fig:beta_CEGC} and \ref{fig:beta_posteriors}). 
Even when including our full array of formation channels, differing assumptions for physical prescriptions alter the recovered branching fractions (see Section~\ref{subsec:fivechannel}). 

While we consider more formation channels in this analysis than has been done before, the astrophysical models used in this work only comprise a subset of the proposed formation channels for \acp{BBH},\footnote{We welcome additional models that could be included in this framework.} and each channel is subject to a number of additional theoretical uncertainties that are not accounted for in this work. 
Therefore, as with any such model selection endeavor that is reliant on population modeling predictions, there are a number of inherent caveats associated with assumptions made for uncertain physical prescriptions. 
A few examples of such caveats are as follows: (a) since a massive-star binary in a \ac{CE} phase has never been observed, the modeling of this phase is entirely theoretical, and the $\alpha$--$\lambda$ energy balance formalism with a fixed \ac{CE} efficiency \alphaCE across all stellar regimes may not be valid; (b) \ac{BH} natal kick magnitudes and orientations relative to the spin axis of the exploding star are uncertain due to the limited observational sample of \ac{BH} binaries with well-measured proper motions, and choosing a natal kick prescription other than the standard bimodal Maxwellian distribution with fallback-modulated kicks can affect both population properties and rates;
(c) using a fixed natal spin for isolated \acp{BH}, which is a proxy for the efficiency of angular momentum transport, may instead be better described by a distribution of natal spins dependent on the properties of the collapsing star; 
and (d) in clusters, changes in the assumed binarity of the primordial population, cluster rotation, triaxiality, and the dynamical effect that would be caused by the presence of a massive \ac{BH} may all have an impact on the properties of \ac{BBH} mergers. 
This list of caveats is not exhaustive, but it provides a sense of the complexity of this model selection problem, especially when considering contributions from multiple formation scenarios that have both shared and independent physical uncertainties. 
Though our quantitative results may change with the inclusion of additional formation models or updated prescriptions for binary stellar evolution and compact-object formation, given the diversity of \ac{BBH} detections to date we anticipate that the necessity for multiple channels significantly contributing to the detected \ac{BBH} population is robust. 
Since the local \ac{BBH} merger rate continues to become more constrained as the catalog of \acp{BBH} grows, predicted merger rates will also be crucial to include in these types of analyses and can be incorporated into branching fraction priors. 

As statistical uncertainties get smaller, systematics become more important~\citep{Barrett2018}, and failing to consider a more complete and comprehensive picture for the diversity of possible \ac{BBH} formation channels will become increasingly dangerous. 
As the detected population of \acp{BBH} grows, our methodology can be expanded to include additional formation channels and uncertain physical prescriptions, which will lead to a more unbiased and complete understanding of the relative contribution from various astrophysical channels to the observed population of compact binary mergers. 
Only through these comprehensive analyses will we be able to accurately infer crucial aspects of \ac{BBH} origins. 

The population models and posterior samples from the hierarchical inference in this work are available on Zenodo~\citep{GWTC2_model_selection_dataset}. 
The codebase developed for this analysis, Astrophysical Model Analysis and Evidence Evaluation (\texttt{AMA}$\mathcal{Z}$\texttt{E}), is available on Github\footnote{\url{https://github.com/michaelzevin/model_selection}} along with notebooks for generating the numbers and figures in this paper.

\acknowledgments
The authors would like to thank Tom Callister and Maya Fishbach for their useful suggestions and to acknowledge the thoughtful comments from the anonymous referees, which helped improve this paper. 
Support for this work and for M.Z. was provided by NASA through the NASA Hubble Fellowship grant HST-HF2-51474.001-A awarded by the Space Telescope Science Institute, which is operated by the Association of Universities for Research in Astronomy, Inc., for NASA, under contract NAS5-26555. 
S.S.B. and T.F. are supported by a Swiss National Science Foundation professorship grant (project No. PP00P2 176868). This project has received funding from the European Union's Horizon 2020 research and innovation program under the Marie Sklodowska-Curie RISE action, grant agreement No.\ 691164 (ASTROSTAT). 
C.P.L.B. is supported by the CIERA Board of Visitors Research Professorship and \ac{NSF} grant PHY-1912648.
V.K. is supported by a CIFAR G+EU Senior Fellowship and Northwestern University. 
P.M. acknowledges support from the FWO junior postdoctoral fellowship No.\ 12ZY520N. 
F.A. acknowledges support from a Rutherford fellowship (ST/P00492X/1) from the Science and Technology Facilities Council. 
D.E.H. was supported by NSF grants PHY-1708081 and PHY-2011997, the Kavli Institute for Cosmological Physics at the University of Chicago, and an endowment from the Kavli Foundation and also gratefully acknowledges a Marion and Stuart Rice Award.
This work used computing resources at CIERA funded by \ac{NSF} grant No.\ PHY-1726951 and resources and staff provided by the Quest high-performance computing facility at Northwestern University, which is jointly supported by the Office of the Provost, the Office for Research, and Northwestern University Information Technology. 
This document has been assigned LIGO document No. \href{https://dcc.ligo.org/LIGO-P2000468/public}{LIGO-P2000468}.

\vspace{5mm}
\software{\texttt{Astropy}~\citep{TheAstropyCollaboration2013,TheAstropyCollaboration2018}; 
\texttt{emcee}~\citep{Foreman-Mackey2013}; 
\texttt{iPython}~\citep{ipython}; 
\texttt{Matplotlib}~\citep{matplotlib}; 
\texttt{NumPy}~\citep{numpy,numpy2}; 
\texttt{Pandas}~\citep{pandas}; 
\texttt{PyCBC}~\citep{PyCBC_v1.14.4}; 
\texttt{SciPy}~\citep{scipy};
\texttt{AMA}$\mathcal{Z}$\texttt{E}~(this work).}

\clearpage
\appendix

\section{Population Models}\label{app:pop_models}

In this section, we provide further details for the population modeling in this work. 
In Appendix~\ref{app:formation_channels}, we discuss the formation channels used in our inference and the assumptions that were made to provide more self-consistency between models. 
In Appendix~\ref{app:star_formation_history}, we discuss the distribution of systems across cosmic time for our five formation channels, as well as the assumed distribution of metallicities as a function of redshift. 

\subsection{Formation Channels}\label{app:formation_channels}

\subsubsection{Isolated Evolution through CE and Stable Mass Transfer}\label{app:CESMT}

The \texttt{CE} and \texttt{SMT} models are simulated with the \texttt{POSYDON} framework (T. Fragos et al. 2021, in preparation), which was used to combine the rapid population synthesis code \texttt{COSMIC}~\citep{Breivik2020} with \texttt{MESA} detailed binary evolution calculations \citep{Paxton2011,Paxton2013,Paxton2015,Paxton2018,Paxton2019} as in \citet{Bavera2020b}. 
\texttt{COSMIC} was used to rapidly evolve binaries from the zero-age main sequence until the end of the second mass-transfer episode. 
For the last phase of the binary evolution (\ac{BH}--Wolf--Rayet), which determines the second-born \ac{BH} spin \citep{Qin2018,Bavera2020}, we used detailed stellar and binary simulations from \texttt{MESA}. 
These simulations take into account differential stellar rotation, tidal interaction, stellar winds, and the evolution of the Wolf--Rayet stellar structure, therefore allowing us to carefully model the tidal spin-up phase until the core collapse of the secondary.

These models assume the first-born \ac{BH} is formed with a negligible spin $\chi_\mathrm{b} \simeq 0$ because of the assumed efficient angular momentum transport~\citep{Qin2018,Fuller2019b} and the Eddington-limited accretion efficiency onto compact objects; this also leads to small first-born \ac{BH} spins for the \texttt{SMT} channel because the mass accreted onto \acp{BH} during the second mass- transfer is negligible \citep{Thorne1974}. 
In this work we artificially varied the first assumption by changing the birth spins in post-processing. 
Assumptions for the efficiency of accretion onto \acp{BH} may affect the natal spins of the first-born \ac{BH} in the \texttt{SMT} channel if it is highly super-Eddington, though \cite{Bavera2020b} showed that a highly super-Eddington accretion efficiency leads to the extinction of \ac{BBH} mergers in the \texttt{SMT} channel and thus we do not consider variations in its value. 

These simulations were designed as much as possible to match the same stellar and binary physical assumptions made in the \texttt{CHE} models; in fact, the \texttt{MESA} model is entirely self-consistent with that used in \citet{duBuisson2020}. 
Consistency in the initial binary distributions was also a priority. 
For example, we assumed that log-initial orbital period distributions follow a \cite{Sana2012} power law in the range $[10^{0.15},10^{5.5}]$~days and extend down to $0.4$~days assuming a flat-in-log distribution in order to sample the parameter space leading to a chemically homogeneous evolution~\citep{duBuisson2020}. 
Finally, analogous to the \texttt{CHE} and \texttt{NSC} models, we used the same prescriptions for distributing the synthetic \ac{BBH} populations across cosmic history (see Appendix~\ref{app:star_formation_history}). 
To translate the underlying \ac{BBH} population to the detected population in all channels, \cite{Bavera2020b} assumed the detection probabilities detailed in Appendix~\ref{app:detection_probabilities} but with a higher network \ac{SNR} threshold of $\rho_{\rm thresh}=12$. 
The estimated rate densities for the \texttt{CE} channel are in the range $17$--$113~\mathrm{Gpc}^{-3}\,\mathrm{yr}^{-1}$ depending on $\alpha_\mathrm{CE}$ (the smallest value corresponds to the model with $\alpha_\mathrm{CE} = 5.0$, while the largest value corresponds to $\alpha_\mathrm{CE}=0.2$), and $25~\mathrm{Gpc}^{-3}\,\mathrm{yr}^{-1}$ for the \texttt{SMT} channel~\citep{Bavera2020b}. 
For a detector network with \texttt{midhighlatelow} sensitivity and a network detection threshold of $\rho_{\rm thresh}=12$, these values translate to a detection rate of $15$--$412~\mathrm{yr}^{-1}$ and $86~\mathrm{yr}^{-1}$ for \texttt{CE} and \texttt{SMT}, respectively.

\subsubsection{Chemically Homogeneous Evolution}\label{app:CHE}

The \texttt{CHE} models are adopted from \cite{duBuisson2020} who computed a large grid of detailed \texttt{MESA} stellar and binary simulations undergoing this evolutionary process. 
For consistency with the other models, we restrict primary masses to the range $[0.01,150]\,\mathrm{M}_\odot$, meaning that, compared to the original study, we ignore systems forming \acp{BBH} with components above the pair instability mass gap~\citep[e.g.,][]{Woosley2017,Farmer2019,Marchant2019}.
The core collapse of the stars' profiles is done self-consistently in \texttt{CE} and \texttt{SMT} models using \texttt{POSYDON} \citep[see Appendix D of][]{Bavera2020b}. 
The applied prescription takes into account disk formation during the collapse of highly spinning stars, mass loss through neutrinos, (pulsational) pair instability supernovae according to the fits of the detailed simulation of \cite{Marchant2019}, and two Blaauw kicks~\citep{Blaauw1961,Kalogera1996} where we assume circularization after the first supernova \citep[see][]{duBuisson2020}. 
The synthetic population of \acp{BBH} is distributed across the Universe cosmic history assuming the same initial binary distributions as in the \texttt{CE} and \texttt{SMT} models. 

Since we have a regular grid of \texttt{MESA} simulations covering the initial binary distributions instead of sampling them with a Monte Carlo approach, we can directly calculate their phase space volume. 
Given a binary $k$ with initial primary mass 
$m_{1,k}$, mass ratio $q_{k}$, and period $p_{k}$, the relative contribution of that system to the total population $P_{k}$ is 
\begin{equation}
    P_{k} \equiv p(m_{1,k}, q_{k}, p_{k}) =
    p_\mathrm{IMF}(m_{1,k}, q_{k}, p_{k}) \times
    p_\mathrm{IQF}(m_{1,k}, q_{k}, p_{k}) \times
    p_\mathrm{IPF}(m_{1,k}, q_{k}, p_{k}) ,
\end{equation}
where $\mathrm{IMF}$, $\mathrm{IQF}$, and $\mathrm{IPF}$ designate the initial mass, mass ratio, and period functions. 
These probabilities are obtained by integrating the assumed initial distribution probability densities independently; for simplicity, we assume that the initial binary properties are independent of each other and of metallicity. 
For the initial mass function, we assume a \citet{Kroupa2001} power law in the range $[0.01,150]\,\mathrm{M}_\odot$; for the initial mass ratios, we assume a flat distribution in the range $[0,1]$; and for the initial periods, we assume an extended \citet{Sana2012} log-power law as in Eq.~(A.1) of \citet{Bavera2020b}. 
For each binary the integration is performed around the initial values $m_{1,k}$, $q_{k}$, and $p_{k}$ assuming a volume corresponding to the grid's resolution, namely, $\Delta \log_{10}(m_{1,k} / \Msun) = 0.025$, $\Delta q_{k} = 0.2$, and $\Delta (p_{k}/\mathrm{day}) = 0.025$. 
Even though the simulations for the \texttt{CHE} channel were carried out at a fixed mass ratio value of $q=1$, here we assume that they are representative of resultant \ac{BBH} mass ratios within $[0.8,1]$, and artificially smear the \ac{BBH} mass ratios uniformly across this range while keeping the total mass of the binary fixed. 
This assumption is justified by the findings of \citet{Marchant2016}.

Similar to Eq.~(7) of \citet{Bavera2020b}, the \ac{BBH} merger rate density is calculated in finite-time bins of $\Delta t_i = 100~\mathrm{Myr}$ and log-metallicity bins $\Delta Z_j$ where each binary $k$ is placed at the center of each time bin corresponding to the redshift of formation $z_{\mathrm{f},i}$ and merging at $z_{\mathrm{m},i,k}$. 
Therefore,
\begin{equation}
    R_\mathrm{BBHs} (z_i) = \sum_{\Delta Z_j} \sum_{k} 
     P_{i,j,k} \, f_\mathrm{bin} \frac{\mathrm{fSFR}(z_{\mathrm{f},i})}{\bar{m}_{\star}} \,
    \frac{4 \pi c \, D^2_\mathrm{c}(z_{\mathrm{m},i,k})}{\Delta V_\mathrm{c}(z_i)} \, \Delta t_i~\mathrm{Gpc^{-3}\, yr^{-1}},
    \label{eq:RBBHs}
\end{equation}
where $f_\mathrm{bin}=0.7$ \citep{Sana2012} is the binary fraction, $\bar{m}_{\star}=0.518\,\Msun$ is the average system mass \citep[computed as in Eq.~(A.2) of][]{Bavera2020}, $\mathrm{fSFR}$ is the \ac{SFR} per metallicity range $\Delta Z_j$, $D_\mathrm{c}(z)$ is the comoving distance, and $\Delta V_\mathrm{c}$ is the comoving volume corresponding to $\Delta t_i$. 
Using a network detection threshold of $\rho_{\rm thresh}=12$, we find a merger rate density of $32.9~\mathrm{Gpc}^{-3}\,\mathrm{yr}^{-1}$ and a detection rate for a detector network with \texttt{midhighlatelow} sensitivity of $360~\mathrm{yr}^{-1}$. 
The rate found here is higher than the one found by \citet{duBuisson2020}, $5.8~\mathrm{Gpc}^{-3}\,\mathrm{yr}^{-1}$, for two reasons: (i) The original study assumed a flat-in-log orbital period distribution over the range $[0.4,365.25]$~days compared to the extended log-power law assumed here for consistency with the \texttt{CE} and \texttt{SMT} channels; when we assume the original distribution over the range $[0.4,10^{5.5}]$~days, the rate density decreases to $10.6~\mathrm{Gpc}^{-3}\,\mathrm{yr}^{-1}$. (ii) \citet{duBuisson2020} assumed an \ac{SFR} and metallicity distribution from the cosmological simulations of \cite{Taylor2015}, which predicts less stellar mass formed at low metallicities compared to \cite{Madau2017a}, assuming the metallicities follow a truncated log-normal distribution around the empirical mean of \cite{Madau2017a} and a standard deviation of $0.5~\mathrm{dex}$.

\subsubsection{Globular Clusters}\label{app:GC}

The \ac{GC} models are simulated using the H\'enon-style cluster Monte Carlo code \texttt{CMC}~\citep{Henon1971,Henon1971a,Joshi2000,Pattabiraman2013}. 
\texttt{CMC} has been shown to reproduce both the global cluster properties and the \ac{BBH} populations found in direct $N$-body cluster models in a fraction of the time~\citep{Rodriguez2016}. 
Each cluster model contains all of the necessary physics to describe the dynamical formation of \acp{BBH}. 
Each star and binary in the cluster is evolved with the \ac{BSE} package of \citet{Hurley2000,Hurley2002} with updated prescriptions for stellar winds, compact-object masses, supernova natal kicks, and pulsational-pair instability physics consistent with \texttt{COSMIC}~\citep[][and references therein]{Chatterjee2010,Rodriguez2016a,Rodriguez2018b}. 
The three-body interactions between single stars that produce many \acp{BBH} are treated probabilistically using prescriptions from \citet{Morscher2013}, which have been well tuned to direct $N$-body integrations. 
Furthermore, stars and binaries are allowed to interact through strong three- and four-body encounters, whose outcomes are directly integrated with Fewbody~\citep{Fregeau2007}, a small-$N$ dynamical integrator with relativistic corrections~\citep{Antognini2014,Amaro-Seoane2016,Rodriguez2018b}. 
\acp{BBH} that merge inside the cluster, either as isolated systems or due to prompt \ac{GW} emission during three-body encounters, are given new masses, spins, and \ac{GW} recoil velocities taken from numerical relativity-based fitting formulae~\citep[][Appendix A]{Rodriguez2018b}. 

As the natal spins of \acp{BH} are set at the start of the simulations, no post-processing is necessary across our \chib models. 
We do not consider differing \alphaCE values in our \texttt{GC} models for two reasons: (i) Most \acp{BBH} that go on to merge are processed dynamically and go through partner swaps throughout their evolution in the cluster and do not merge with their original partner; thus the post-\ac{CE} separation has a minimal impact on the rates and properties of \ac{BBH} mergers. (ii) Of the \acp{BH} in the cluster originally in a \ac{BBH} system that formed from a massive-star binary progenitor, we find only a percent-level number that were at tight enough orbital configurations during the \ac{BH}--Wolf--Rayet phase for tidal spin-up to be relevant. 
We therefore set the \ac{CE} efficiency to our fiducial value of $\alphaCE = 1$ in the \texttt{GC} model. 

The \texttt{GC} model is the only model that does not follow the standard star formation and metallicity evolution described in Appendix~\ref{app:star_formation_history}, since cluster formation does not mimic the star formation history of the host galaxies. 
We instead follow the prescriptions in \cite{Rodriguez2018a}, which rely on detailed modeling of \ac{GC} formation across cosmic time~\citep{ElBadry2019}, and weight \acp{GC} of differing metallicities by the metallicity distribution of \acp{GC} observed in the Milky Way~\citep{Harris2010}.

\subsubsection{Nuclear Star Clusters}\label{app:NSC}

The evolution of \texttt{NSC} models is determined using the semi-analytical approach of \citet{Antonini2019a}. 
In this method, we assume that the energy generated by the \ac{BH} binaries in the cluster core is regulated by the process of two-body relaxation in the bulk of the system~\citep{Breen2013a}. 
This principle of balanced evolution~\citep{Henon1961} is used to compute the hardening and the merger rate of the core binaries. 
Moreover, we neglect mass loss from stellar evolution and the escape of \acp{BH} and stars, i.e.\ we assume a constant cluster mass (see \citealt{Antonini2020a} for caveats in this assumption). 
Each \ac{BBH} formed dynamically in the cluster core is then evolved until it either merges inside the cluster or it is ejected from it. 
If the merger remnant is retained inside the cluster, we compute its spin and mass using the prescriptions in \citet{Rezzolla2008}. 
We evolve the cluster until either all \acp{BH} have been ejected or a time of $13~\mathrm{Gyr}$ has passed. 

As with the \texttt{GC} model, we do not consider differing \alphaCE values for the \texttt{NSC} model and assume the population properties are the same across all values of \alphaCE (see discussion in Appendix~\ref{app:GC}). 
In contrast to those in the \texttt{GC} model, we assume that star formation and metallicity evolution follow the same prescriptions as the \texttt{CE}, \texttt{CHE}, and \texttt{SMT} models (i.e.\ they trace the evolution of the host galaxy as a whole). There are arguments that the star formation histories of nuclear clusters are different from those of their galactic hosts \citep{Neumayer2020}. 
Given the uncertainties, however, we continue with the assumption above and
sample from the three metallicity models ($0.01\Zsun$, $0.1\Zsun$, and $1 \Zsun$) according to the prescriptions described in Appendix~\ref{app:star_formation_history}.

\subsection{Formation Rate and Metallicity Evolution}\label{app:star_formation_history}

All formation channels provide raw samples of \ac{BBH} mergers for a given \chib, \alphaCE, and metallicity. 
For all models other than \texttt{GC} (see Appendix~\ref{app:GC}), we distribute the synthetic \ac{BBH} populations across cosmic history assuming the \ac{SFR} density in \cite{Madau2017a}: 
\begin{equation}\label{eq:sfr_density}
    \psi(z) = 10^{-2} \frac{(1+z)^{2.6}}{1 + \left[(1+z)/3.2\right]^{6.2}}\,\Msun\,\mathrm{yr}^{-1}\,\mathrm{Mpc}^{-3}. 
\end{equation}
This determines the birth redshift of the \ac{BBH} progenitor. 
For the \texttt{CE}, \texttt{CHE}, and \texttt{SMT} models, the merger redshift is then calculated using the \ac{BBH} formation time ($t_\mathrm{birth} - t_\mathrm{BBH}$) and inspiral time ($t_\mathrm{insp}$), the latter of which is determined using the orbital properties of the binary following the birth of the second \ac{BH}~\citep{Peters1964}. 
Thus, the merger redshift is 
\begin{equation}
    z_\mathrm{merge} = \mathcal{T}(t_\mathrm{birth}-t_\mathrm{BBH}-t_\mathrm{insp}),
\end{equation}
where $\mathcal{T}$ is the transformation function between the lookback time and redshift. 
For the \texttt{NSC} model, delay times ($t_\mathrm{delay} = t_\mathrm{BBH}+t_\mathrm{insp}$) are computed directly from the model and used to determine the merger redshift. 
For all models, we assume a $\Lambda$CDM cosmology with the \textit{Planck 2015} cosmological parameters of ${H_0 = 68~\mathrm{km\,s}^{-1}\,\mathrm{Mpc}^{-1}}$, ${\Omega_\mathrm{m} = 0.31}$, and ${\Omega_{\Lambda} = 0.69}$~\citep{PlanckCollaboration2016}. 

Each formation channel model is simulated across a range of metallicities. 
At a given redshift, metallicities are distributed following a truncated log-normal metallicity distribution around the empirical median metallicity from \cite{Madau2017a} assuming a standard deviation of $0.5~\mathrm{dex}$ \citep[][Section~2.2]{Bavera2020b}: 
\begin{equation}\label{eq:metallicity}
    \log_{10}\left\langle Z/\Zsun \right\rangle  = 0.153 - 0.074 z^{1.34},
\end{equation}
with a solar metallicity of ${\Zsun=0.017}$~\citep{Grevesse1998}. 
We use the \ac{SFR} density, Eq.~\eqref{eq:sfr_density}, and metallicity distribution, Eq.~\eqref{eq:metallicity}, to construct a full cosmological population for each submodel of the formation channels (parameterized by \chib and \alphaCE).

\section{Detection Probabilities}\label{app:detection_probabilities}

In our inference, detection probabilities are a key component of the detection efficiency $\xi$ in the hyperlikelihood. 
From the cosmological populations of each channel, we calculate detection probabilities numerically. 
Though this is more computationally intensive than using analytical scaling relations that approximate the sensitive spacetime volume to leading order~\citep[e.g.,][]{Fishbach2017a}, we choose to calculate detection probabilities numerically to better capture the influence that total mass, mass ratio, and spins have on selection effects. 

Each system is characterized by its (source-frame) component masses, three-dimensional component spin vectors, and merger redshift. 
For every system in each population model, we first calculate the optimal \ac{SNR} $\rho_\mathrm{opt}$ for LIGO Hanford, LIGO Livingston and Virgo operating at \texttt{midhighlatelow} sensitivity~\citep{LVC_ObservingScenarios} by assuming the system is directly overhead with a face-on inclination. 
We use the \texttt{IMRPhenomPv2} waveform approximant \citep{Hannam2014,Khan2016} for determining \acp{SNR}, and detector response functions are constructed using the \texttt{PyCBC} package \citep{PyCBC_v1.14.4}. 
We approximate the optimal network \ac{SNR} as the quadrature sum of the optimal \acp{SNR} from the three detectors, 
\begin{equation}
    \rho_\mathrm{net,\,opt}\, \lesssim\, \sqrt{\sum_i (\rho_\mathrm{i,\,opt}^2)}, 
\end{equation}
which will give us a conservative overestimate of the true optimal \ac{SNR} of the network. 
We choose a network \ac{SNR} threshold of $\rho_\mathrm{thresh} = 10$, consistent with the false-alarm-rate threshold of two per year, which is used as a criterion for events in {GWTC-2}~\citep{LVC_ObservingScenarios,Nitz2020b,GWTC2}. 
If $\rho_\mathrm{net,\,opt} < \rho_\mathrm{thresh}$, we set the detection probability of the system to $\tilde{P}_\mathrm{det} = 0$. 
Otherwise, we consider the source potentially detectable and perform $10^3$ Monte Carlo realizations of the extrinsic parameters, namely the right ascension, declination, inclination, and polarization angle. 
The detection probability of the system marginalized over the extrinsic parameters is then given by 
\begin{equation}
    \tilde{P}_{\mathrm{det}} =  \frac{1}{N} \sum^{N}_{j=1} \mathcal{H}\left[ \sqrt{\sum_i (\rho_i(\psi_j))^2} - \rho_\mathrm{thresh} \right],
\end{equation}
where $N$ is the number of Monte Carlo realizations,  $i$ indexes over detectors, $\psi_j$ denotes the extrinsic parameters drawn for the Monte Carlo sample $j$, and $\mathcal{H}$ is the Heaviside step function. 
These detection weights are used to construct the weighted \ac{KDE} models in Figure~\ref{fig:pop_models}.

\section{KDEs of Models}\label{app:kdes}

We use an adaptation of the \texttt{gaussian\_kde} class of \texttt{SciPy} to construct \acp{KDE} for each population model, which are 4-dimensional over the parameters $\vec{\theta}$. 
Our \texttt{gaussian\_kde} class handles reflection over physical boundaries in the parameter space (i.e.\ $0 < q \leq 1$). 
To ensure an adequate choice of \ac{KDE} bandwidth for our population models, we perform a holdout analysis where we construct the \ac{KDE} using a subset of samples from the full population, draw samples from the \ac{KDE}, and compare the one-dimensional marginalizations of the parameters $\vec{\theta}$ drawn from the \ac{KDE} with another subset of samples. 
We find a bandwidth of $\approx\,0.01$ consistently matches the true distribution of parameters, whereas values lower and higher tend to overfit and underfit the data, respectively.

\section{Population Inference}\label{app:inference}

Our goal is to recover the posterior for our set of hyperparameters, ${\vec{\beta} = [\beta_{\texttt{CE}}, \beta_{\texttt{CHE}}, \beta_{\texttt{GC}}, \beta_{\texttt{NSC}}, \beta_{\texttt{SMT}}]}$ and $\vec{\lambda} = [\chib, \alphaCE]$, given the set of (independent) \ac{GW} observations of \acp{BBH} from GWTC-2, $\mathbf{x} = \{\vec{x}_{i}\}_{i}^{N_{\mathrm{obs}}}$. 
In the following, we are only interested in the shape of the populations and not the rate and implicitly marginalize out the rate term by assuming a $p(N) \propto 1/N$ prior on the number of detections~\citep{Fishbach2018}. 

Starting from the ground up, the probability of detecting a set of event parameters $\boldsymbol{\theta} = \{\vec{\theta}_i\}$ given the model hyperparameters $\vec{\Lambda} = [\vec{\beta}, \vec{\lambda}]$ from independent observations is
\begin{equation}\label{eq:p_theta_lambda}
    p(\boldsymbol{\theta} | \vec{\Lambda}) = 
    \prod_{i=1}^{N_\mathrm{obs}} 
    \frac{p(\vec{\theta_i} | \vec{\Lambda})}
    {\int p(\vec{\theta} | \vec{\Lambda})P_\mathrm{det}(\vec{\theta}) \,\mathrm{d}\vec{\theta}}, 
\end{equation}
where $P_\mathrm{det}(\vec{\theta})$ is the detection probability for an event with parameters $\vec{\theta}$~\citep{Chennamangalam2013,Farr2015a,Mandel2019,Vitale2020a}. 
Marginalizing over event parameters, the probability of observing the data for event $\vec{x}_i$ given our hyperparameters $\vec{\Lambda}$ is 
\begin{equation}
    p(\vec{x}_i | \vec{\Lambda}) = 
    \int p(\vec{x}_i | \vec{\theta}) p(\vec{\theta} | \vec{\Lambda}) \,\mathrm{d}\vec{\theta} . 
\end{equation}
By applying Bayes' theorem, we replace $p(\vec{x}_i | \vec{\theta})$ with $p(\vec{\theta} | \vec{x}_i) p(\vec{x}_i) / \pi(\vec{\theta})$. 
Assuming independent observations, we get
\begin{equation}
    p(\mathbf{x} | \vec{\Lambda}) = 
    \prod_{i=1}^{N_\mathrm{obs}} \frac{p(\vec{x}_i)}{\int p(\vec{\theta} | \vec{\Lambda})P_\mathrm{det}(\vec{\theta}) \,\mathrm{d}\vec{\theta}} \int
    \frac{p(\vec{\theta}_i | \vec{x}_i) p(\vec{\theta}_i | \vec{\Lambda})}
    {\pi(\vec{\theta}_i)} \,\mathrm{d}\vec{\theta} ,
\end{equation}
where $\pi(\vec{\theta})$ is the prior on the parameters $\vec{\theta} = [\Mc, q, \chi_\mathrm{eff}, z]$ assumed in the original inference of $\vec{\theta}$, which is provided alongside the \ac{LVC} posterior samples. 
We evaluate $\pi(\vec{\theta})$ at each point $\vec{\theta}_i$ in a 4-dimensional prior \ac{KDE} constructed using the \ac{LVC} prior samples. 
Since we use $S_i$ posterior samples to approximate $p(\vec{\theta}_i | \vec{x}_i)$, we can rewrite this integral as a discrete sum over the posterior samples: 
\begin{equation}\label{eq:app_hyperlikelihood1}
    p(\mathbf{x} | \vec{\Lambda}) = 
    \prod_{i=1}^{N_\mathrm{obs}} \frac{p(\vec{x}_i)}{S_i \int p(\vec{\theta} | \vec{\Lambda})P_\mathrm{det}(\vec{\theta}) \,\mathrm{d}\vec{\theta}} \sum_{k=1}^{S_i} 
    \frac{p(\vec{\theta}_i^k | \vec{\Lambda})}{\pi(\vec{\theta}_i^k)} .
\end{equation}

The hyperlikelihood $p(\vec{\theta}_i^k | \vec{\Lambda})$ is evaluated as a mixture model of the underlying \acp{KDE} in the current \chib and \alphaCE model, 
\begin{equation}
    p(\vec{\theta}_i^k | \vec{\Lambda}) = \sum_j \beta_j p(\vec{\theta}_i^k | \mu_j^{\chi, \alpha}), 
\end{equation}
where the summation is over the formation channels and $\mu_j^{\chi, \alpha}$ is the \chib and \alphaCE model that the sampler is in at a given step in the chain. 
Thus, our hyperlikelihood from Eq.~\eqref{eq:app_hyperlikelihood1} becomes 
\begin{equation}
    p(\mathbf{x} | \vec{\Lambda}) = 
    \prod_{i=1}^{N_\mathrm{obs}} \frac{p(\vec{x}_i)}{S_i \tilde{\xi}^{\chi, \alpha}}
    \sum_{j} \beta_j
    \sum_{k=1}^{S_i} \frac{p(\vec{\theta}_i^k | \mu_j^{\chi, \alpha})}{\pi(\vec{\theta}_i^k)}, 
\end{equation}
where for convenience we define $\tilde{\xi}^{\chi, \alpha} \equiv \sum_j \beta_j \xi_j^{\chi, \alpha}$ where 
\begin{equation}
\xi_j^{\chi, \alpha} = \int p(\vec{\theta}|\mu_j^{\chi, \alpha})P_\mathrm{det}(\vec{\theta})\, \mathrm{d}\vec{\theta}
\end{equation}
is the detection efficiency for each formation channel model with natal spin $\chi$ and \ac{CE} efficiency $\alpha$. 
The channel-dependent detection efficiency $\xi_j^{\chi, \alpha}$ is evaluated using a Monte Carlo approach, since detection probabilities are already calculated for each sample in the population models. 
Finally, the posterior distribution on the hyperparameters, $p(\vec{\Lambda} | \mathbf{x}) = p(\mathbf{x} | \vec{\Lambda}) \pi(\vec{\Lambda}) / p(\mathbf{x})$, is
\begin{equation}
    p(\vec{\Lambda} | \mathbf{x}) = 
    \pi(\vec{\Lambda}) \prod_{i=1}^{N_\mathrm{obs}} \frac{1}{S_i \tilde{\xi}^{\chi, \alpha}}
    \sum_{j} \beta_j 
    \sum_{k=1}^{S_i} \frac{p(\vec{\theta}_i^k | \mu_j^{\chi, \alpha})}{\pi(\vec{\theta}_i^k)} ,
\end{equation}
where $\pi(\vec{\Lambda})$ is the prior on the hyperparameters. 

For priors, we use a Dirichlet distribution with equal concentration parameters and dimensions equal to the number of formation channels as a prior for the branching fractions $\vec{\beta}$, imposing the constraints $(0 \leq \beta_i \leq 1) \,\forall\, i$ and $\sum_i \beta_i = 1$. 
In practice, the discrete \chib and \alphaCE models are sampled using dummy index parameters that are defined on the range $[0, N_{\lambda,m}]$, where $N_{\lambda,m}$ is the number of $m=\chib$ or $n=\alphaCE$ models, with a flat prior across this range and no support outside this range. 
The $\vec{\lambda}$ model considered at each step is given by the floor of the dummy parameter values that correspond to \chib and \alphaCE.

\section{Testing with Mock Observations}\label{app:mock_obs}

\begin{figure*}[t]
\includegraphics[width=0.98\textwidth]{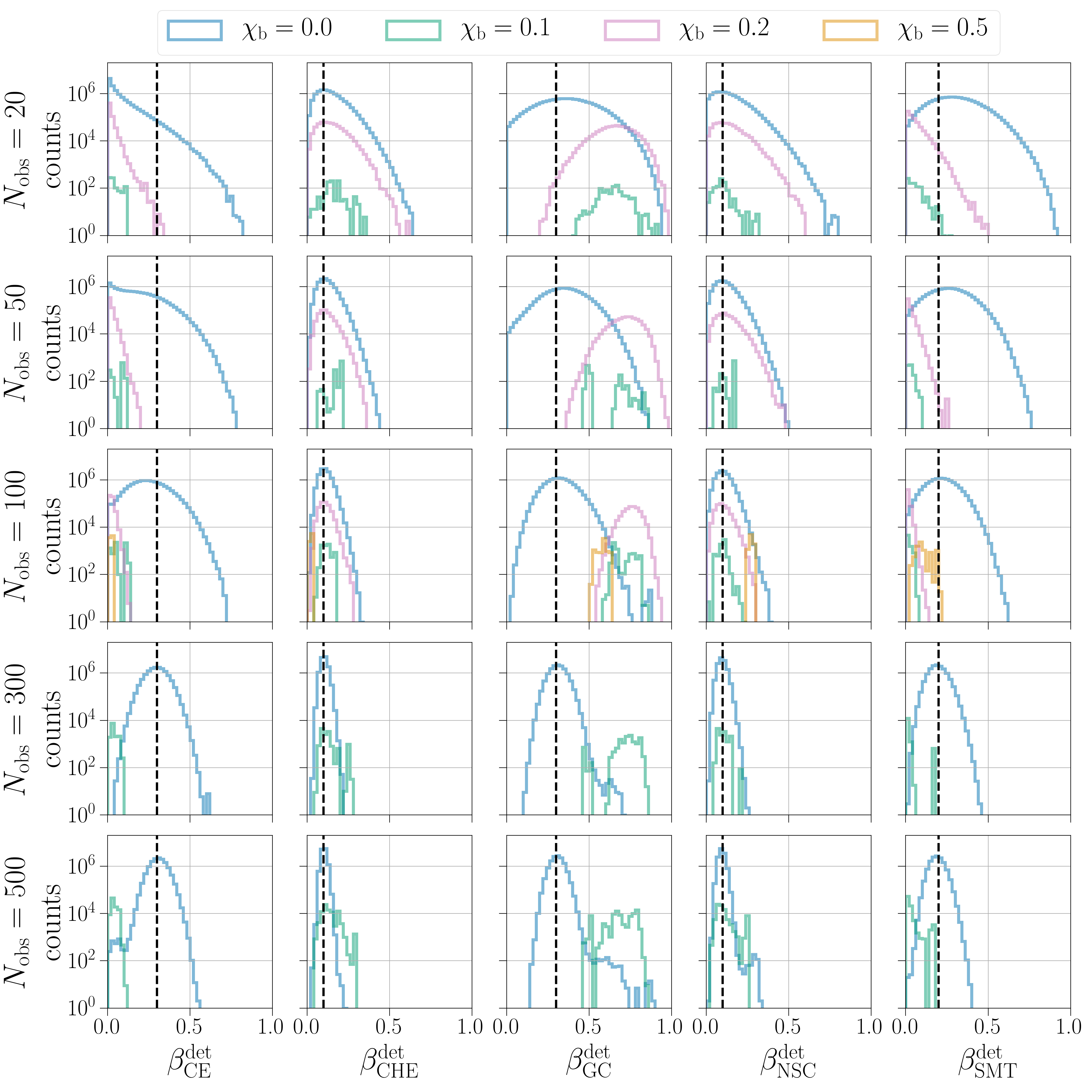}
\caption{Convergence on underlying branching fractions using mock observations from the population models. 
In this example, the injected model has $\chib=0$, $\alphaCE=1.0$, and $\vec{\beta}^\mathrm{det} = [0.3, 0.1, 0.3, 0.1, 0.2]$. 
Different colors show the contribution of different \chib models to the full branching fraction posteriors, marginalized over \alphaCE models; the injected \chib model is shown with the blue curve. 
The Bayes factors between the physical models are the relative areas under the colored curves. 
}
\label{fig:mock_obs_chib}
\end{figure*}

\begin{figure*}[t]
\includegraphics[width=0.98\textwidth]{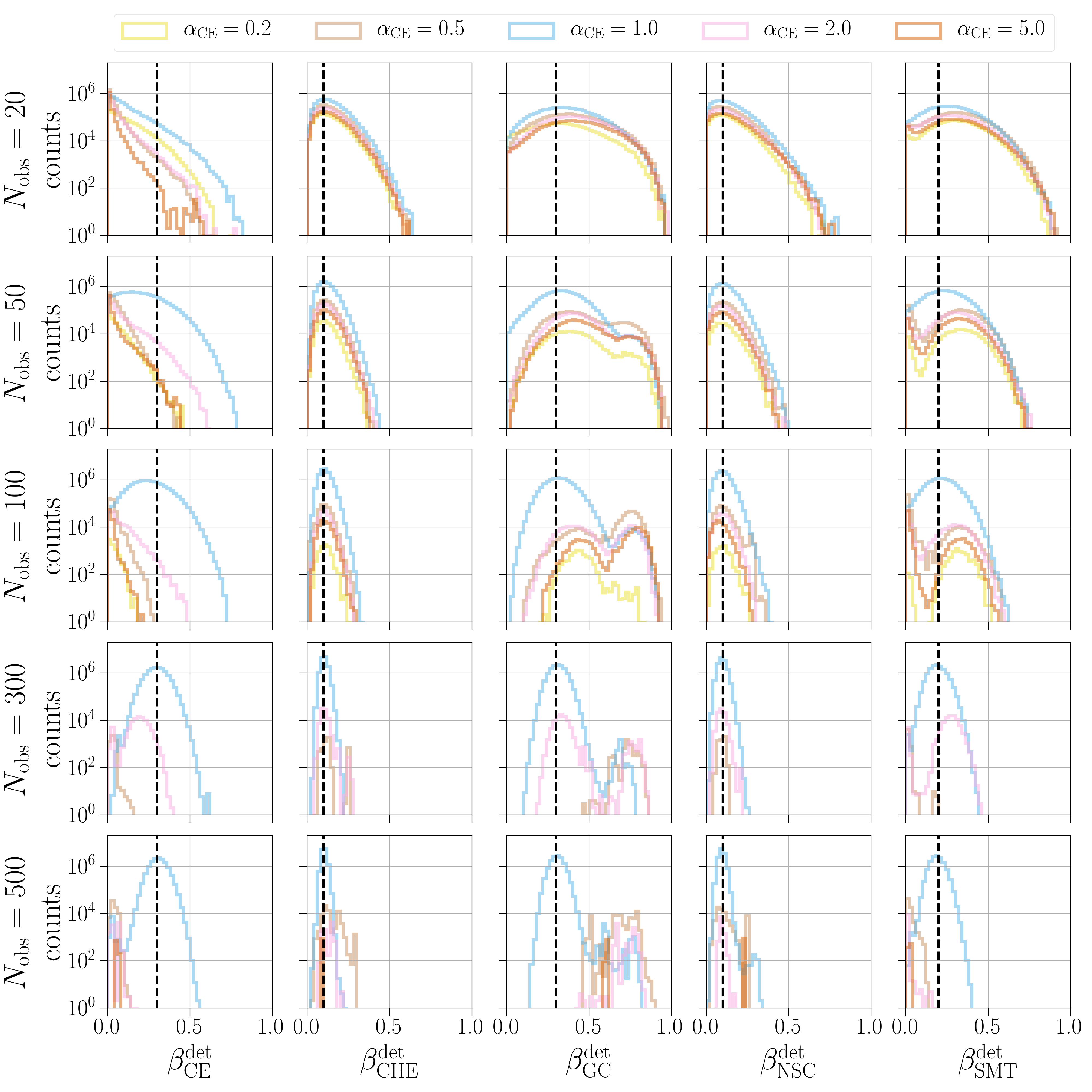}
\caption{Same as Figure~\ref{fig:mock_obs_chib}, but the different colors show the contribution of different \alphaCE models to the full branching fraction posteriors, marginalized over \chib models. 
The injected \alphaCE model is shown with the light blue curve. 
}
\label{fig:mock_obs_alphaCE}
\end{figure*}

In addition to examining constraints on the \ac{GW} population, we can test our methodology using mock draws from the underlying population distributions. 
In Figures~\ref{fig:mock_obs_chib} and \ref{fig:mock_obs_alphaCE}, we show the convergence on detectable branching fractions and physical prescriptions as the number of observations increases. 
In this mock sample, we set the true physical prescriptions to $\chib = 0.0$ and $\alphaCE = 1.0$ and the detectable branching fractions between channels to be $[\beta_\texttt{CE}^\mathrm{det}, \beta_\texttt{CHE}^\mathrm{det}, \beta_\texttt{GC}^\mathrm{det}, \beta_\texttt{NSC}^\mathrm{det}, \beta_\texttt{SMT}^\mathrm{det}] = [0.3, 0.1, 0.3, 0.1, 0.2]$. 
We draw systems from the underlying distributions of the various populations until $N_\mathrm{obs} \beta_j^\mathrm{det}$ detectable samples are drawn from channel $j$, where $N_\mathrm{obs}$ is the number of observed events for a particular mock realization. 
Figure~\ref{fig:mock_obs_chib} shows the contribution to branching fraction posteriors for different \chib models, marginalized over \alphaCE models, and Figure~\ref{fig:mock_obs_alphaCE} shows the contribution to branching fraction posteriors for different \alphaCE models, marginalized over \chib models. 
For demonstration purposes, in these examples we assume no measurement uncertainty; in actuality the inclusion of mock measurement uncertainty will lead to less precise measurements. 
In this simplified example, we find our analysis recovers the injected model, with increasing Bayes factors for the correct physical prescription and increasing precision in the branching fraction measurement as the number of observations increases. 
As with our analysis using the \ac{GW} observations, we see strong biases in the recovered branching fractions when the incorrect physical prescriptions are considered. 
A more investigative analysis with the inclusion of \ac{SNR}-dependent measurement uncertainty will be explored in future work. 

\bibliography{library}{}
\bibliographystyle{aasjournal}

\end{document}